\documentclass[preprint,times]{aastex61}
\setcitestyle{square,numbers}
\usepackage{amsmath} 
\usepackage{amssymb}
\usepackage[OT1]{fontenc}

\renewcommand{\thefootnote}{\Alph{footnote}}

\accepted{to \emph{Nature Astronomy}}

\graphicspath{ {images/} }

\shorttitle{M~dwarfs at low radio frequencies}
\shortauthors{Callingham et al.} 

\begin{document}

\title{The population of M~dwarfs observed at low radio frequencies} 

\correspondingauthor{J.~R.~Callingham}
\email{jcal@strw.leidenuniv.nl}

\author[0000-0002-7167-1819]{J.~R.~Callingham}
\affil{Leiden Observatory, Leiden University, PO\,Box 9513, 2300\,RA, Leiden, The Netherlands}
\affiliation{ASTRON, Netherlands Institute for Radio Astronomy, Oude Hoogeveensedijk 4, Dwingeloo, 7991\,PD, The Netherlands}

\author[0000-0002-0872-181X]{H.~K.~Vedantham}
\affiliation{ASTRON, Netherlands Institute for Radio Astronomy, Oude Hoogeveensedijk 4, Dwingeloo, 7991\,PD, The Netherlands}
\affiliation{Kapteyn Astronomical Institute, University of Groningen, PO\,Box 72, 97200\,AB, Groningen, The Netherlands}

\author[0000-0001-5648-9069]{T.~W.~Shimwell}
\affiliation{ASTRON, Netherlands Institute for Radio Astronomy, Oude Hoogeveensedijk 4, Dwingeloo, 7991\,PD, The Netherlands}
\affiliation{Leiden Observatory, Leiden University, PO\,Box 9513, 2300\,RA, Leiden, The Netherlands}

\author[0000-0003-2595-9114]{B.~J.~S.~Pope}
\affiliation{School of Mathematics and Physics, The University of Queensland, St Lucia, QLD 4072, Australia}
\affiliation{Centre for Astrophysics, University of Southern Queensland, West Street, Toowoomba, QLD 4350, Australia}

\author{I.~E.~Davis}
\affiliation{Cahill Center for Astronomy and Astrophysics, California Institute of Technology, Pasadena, CA 91125, USA}
\affiliation{ASTRON, Netherlands Institute for Radio Astronomy, Oude Hoogeveensedijk 4, Dwingeloo, 7991\,PD, The Netherlands}

\author{P.~N.~Best}
\affiliation{SUPA, Institute for Astronomy, Royal Observatory, Blackford Hill, Edinburgh, EH9 3HJ, UK}

\author{M.~J.~Hardcastle}
\affiliation{Centre for Astrophysics Research, University of Hertfordshire, College Lane, Hatfield AL10 9AB, UK}

\author{H.~J.~A.~R\"{o}ttgering}
\affiliation{Leiden Observatory, Leiden University, PO\,Box 9513, 2300\,RA, Leiden, The Netherlands}

\author{J.~Sabater}
\affiliation{SUPA, Institute for Astronomy, Royal Observatory, Blackford Hill, Edinburgh, EH9 3HJ, UK}

\author{C.~Tasse}
\affiliation{GEPI, Observatoire de Paris, Universit\'{e} PSL, CNRS, 5 Place Jules Janssen, 92190 Meudon, France}
\affiliation{Department of Physics \& Electronics, Rhodes University, PO Box 94, Grahamstown 6140, South Africa}

\author{R.~J.~van Weeren}
\affiliation{Leiden Observatory, Leiden University, PO\,Box 9513, 2300\,RA, Leiden, The Netherlands}

\author{W.~L.~Williams}
\affiliation{Leiden Observatory, Leiden University, PO\,Box 9513, 2300\,RA, Leiden, The Netherlands}

\author{P.~Zarka}
\affiliation{Station de Radioastronomie de Nan\c{c}ay, Observatoire de Paris, PSL Research University, CNRS, Universit\'{e} Orl\'{e}ans, OSUC, 18330 Nan\c{c}ay, France}
\affiliation{LESIA, Observatoire de Paris, CNRS, PSL, Meudon, France}

\author{F.~de Gasperin}
\affiliation{Hamburger Sternwarte, Universit\"{a}t Hamburg, Gojenbergsweg 112, 21029, Hamburg, Germany}

\author{A.~Drabent}
\affiliation{Th\"{u}ringer Landessternwarte, Sternwarte 5, D-07778 Tautenburg, Germany}

\begin{abstract}

\noindent\textbf{Coherent low-frequency ($\lesssim 200$\,MHz) radio emission from stars encodes the conditions of the outer corona, mass-ejection events, and space weather\cite{1985ARA&A..23..169D,2019ApJ...871..214V,2004ApJ...612..511L,2007P&SS...55..598Z,2018ApJ...854...72T}. Previous low-frequency searches for radio emitting stellar systems have lacked the sensitivity to detect the general population, instead largely focusing on targeted studies of anomalously active stars\cite{2019ApJ...871..214V,2016ApJ...830...24C,2018MNRAS.478.2835L,harish2019,2017ApJ...836L..30L}. Here we present 19 detections of coherent radio emission associated with known M~dwarfs from a blind flux-limited low-frequency survey. Our detections show that coherent radio emission is ubiquitous across the M~dwarf main sequence, and that the radio luminosity is independent of known coronal and chromospheric activity indicators. While plasma emission can generate the low-frequency emission from the most chromospherically active stars of our sample\cite{1985ARA&A..23..169D,Gudel2002}, the origin of the radio emission from the most quiescent sources is yet to be ascertained. Large-scale analogues of the magnetospheric processes seen in gas-giant planets\cite{2007P&SS...55..598Z,2004JGRA..109.9S15Z,2018A&A...618A..84Z} likely drive the radio emission associated with these quiescent stars. The slowest-rotating stars of this sample are candidate systems to search for star-planet interaction signatures.}

\end{abstract}

\section*{Main Body} \label{sec:body}

Low-frequency ($\lesssim$\,200\,MHz) radio observations are one of the few tools available that can probe the conditions of the outer coronae of stars\cite{1985ARA&A..23..169D,2019ApJ...871..214V}. For example, magnetic eruptive events, such as coronal mass ejections (CMEs), are often accompanied by low-frequency emission as the ejected plasma traverses the corona into interplanetary space\cite{1950AuSRA...3..387W}. In such situations it is expected that circularly polarised low-frequency emission would be generated by plasma mechanisms in stellar coronae\cite{1985ARA&A..23..169D}, providing a measure of coronal density.

It has also been predicted that we should expect low-frequency radio emission to be generated by a satellite interacting with its host star via the electron-cyclotron maser instability (ECMI)\cite{2007P&SS...55..598Z,2016pmf..rept.....L}. The expected signpost of a star-planet interaction associated with M~dwarfs -- which are significantly less massive but more magnetically-active than our Sun, and are the most common stars in our Milky Way\cite{2015ApJ...806...41C,2018MNRAS.481.5146B,Khodachenko2007,2019NatAs.tmp..408C} -- is emission similar to that produced by magnetospheric processes in gas-giants of the Solar System\cite{2007P&SS...55..598Z}. Such emission is expected to be low-frequency and highly circularly polarised ($\gtrsim$\,60\%), with a brightness temperature exceeding 10$^{12}$\,K\cite{1985ARA&A..23..169D,harish2019,1998JGR...10320159Z}. 

While low-frequency radio emission provides unique information about the magnetic field and environmental conditions around M~dwarfs, particularly about high coronal layers, previous low-frequency blind searches have lacked the required sub-mJy sensitivity necessary to achieve a detection\cite{2018MNRAS.478.2835L}. In addition, prior low-frequency observations of M~dwarfs have focused on the most chromospherically active stars, largely because it has been hypothesised that the radio emission from CMEs might correlate with flare activity\cite{2019ApJ...871..214V,2017ApJ...836L..30L,2018ApJ...862..113C}. Such stars are unique and it is unclear if they are representative of the general low-frequency radio-emitting population. With the LOw Frequency Array (LOFAR), we are currently conducting the LOFAR Two-meter Sky Survey (LoTSS)\cite{2019A&A...622A...1S} -- the deepest, wide-field low-frequency radio survey performed to date. At 144\,MHz, we can achieve an optimal root-mean-square noise level of $\approx$60\,$\mu$Jy per 8\,hr exposure in circular polarisation (Stokes~V)\cite{2019A&A...622A...1S}. With $\approx$20\% of the Northern sky processed, we have examined the LoTSS Stokes~V maps for $\geq4\sigma$ sources and crossmatched those detections to sources in the \emph{Gaia} Data Release 2 (DR2)\cite{2018A&A...616A...1G} to search for radio counterparts to stellar systems\cite{2019RNAAS...3...37C} (see methods). 

\begin{figure}
   \centerline{\resizebox{0.8\textwidth}{!}{\includegraphics{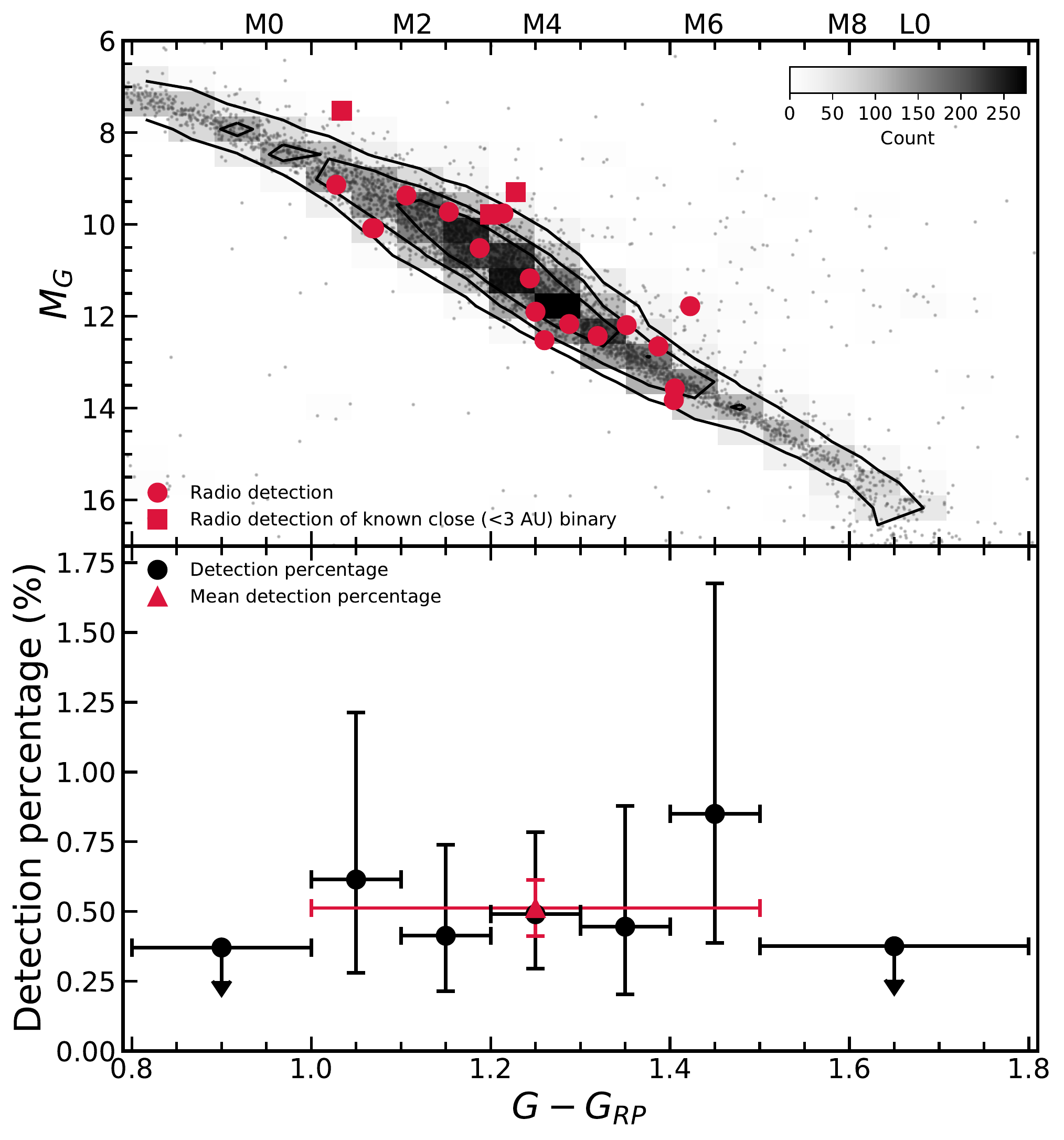}}}
    \caption{\label{fig:ms_dectstats}
    {\bf Our low-frequency radio detections highlighted on a \emph{Gaia} colour ($G - G_{RP}$)-magnitude ($M_{G}$) diagram (top panel), and our low-frequency radio stellar system detection statistics for \emph{Gaia} DR2 (bottom panel).} 
 The \emph{Gaia} source population plotted consists of a clean sample of M and L dwarfs that are within 50\,pc and above a Galactic latitude of 20$^{\circ}$\cite{2019AJ....157..231K} (see methods). In the top panel, the LoTSS detected stellar systems are indicated by red circles, with red squares highlighting known close ($<$\,3\,AU separation) binaries. The colour of the density map represents the number of sources in each coloured pixel, as set by the colour bar in the top-right of the plot. The contour levels correspond to 25, 75, and 150 sources, respectively. In the bottom panel, our LoTSS-\emph{Gaia} DR2 detection statistics for different \emph{Gaia} colour bins are shown as black circles, with the mean of all our detections plotted as a red triangle. Upper-limits and error bars represent 1-$\sigma$ uncertainty.}
\end{figure}

In Table\,\ref{tab:radio_detections}, we present 19 new low-frequency detections of M~dwarfs, with only the LoTSS detection of GJ\,1151 having been presented in a previous publication\cite{harish2019}. The size of the sample represents over an order of magnitude increase in the number of M~dwarfs detected blindly at low frequencies, allowing us to perform the first population study not biased by targeting strategies. As shown in Figure\,\ref{fig:ms_dectstats}, the sample spans the main-sequence spectral range M1.5 to M6.0, implying that we have detected systems with M~dwarfs that are both partially and fully convective. We detect $0.5\pm0.1\%$ of the stars in \emph{Gaia} DR2 that are within 50\,pc and which have spectral types M1.5 to M6.0. Each spectral type has an equal probability of detection, within uncertainty, although we note that the \emph{Gaia} catalogue begins to become significantly incomplete for spectral types later than $\sim$M7\cite{2019AJ....157..231K}. Three of our detections (DG\,CVn, CR\,Dra, and GJ\,3729) are known binaries with less than 3\,AU separation, which is consistent with the known binarity rate of M~dwarfs\cite{2015ApJ...813...75S}.

\begin{table}
\centering
  \caption{\footnotesize \label{tab:radio_detections} {\bf Radio and optical properties of the 144\,MHz detected M~dwarfs.} Sp. Type, $d$, $S_{V}$, $S_{V}/S_{I}$, $M$, $N$, $P_{\mathrm{rot}}$, and $T_{B}$ correspond to spectral type, distance, Stokes~V flux density, percentage of circularly polarised radio light, number of detections, number of exposures, stellar rotation, and 144\,MHz brightness temperature, respectively. Known close ($<$\,3\,AU separation) binaries are identified by an asterisk, and the reported spectral type is for the total system. $T_{B}$ is a lower limit as it was calculated assuming that the emission emanates from the total photospheric surface. Potential ECMI sites are likely confined to much smaller areas, as observed on Jupiter\cite{2004JGRA..109.9S15Z,1998JGR...10320159Z}. Stellar rotation periods are not reported if unknown. Uncertainties on rotational periods are not listed if they are smaller than the last significant figure.}
 \begin{center}
    \begin{tabular}{llcccccc}
      \hline
      Name & Sp.\,Type & $d$ & $S_{V}$ & $|S_{V}/S_{I}|$ & $M$,\,$N$ & $P_{\mathrm{rot}}$ & $T_{B}$ \\
       &  & (pc) & (mJy) & \% & & (days) & ($\times 10^{12}$\,K) \\
       \hline
       \hline 
  DO\,Cep & M4.0 & 4.00 & $-1.83\pm$0.13 & 38$\pm$5 & 1,\,1 & 0.41 &  0.7$\pm$0.1\\
  WX\,UMa & M6.0 & 4.90 & $-1.52\pm$0.15 & 96$\pm$4 & 3,\,3  & 0.78 & 2.6$\pm$0.4\\
  AD\,Leo & M3.0 & 4.97 & $-1.34\pm$0.11 & 41$\pm$10 & 2,\,2 & 2.23 & 0.4$\pm$0.1\\
  GJ\,625 & M1.5 & 6.47 & $-1.20\pm$0.19 & 75$\pm$15 & 6,\,18 & 79.8$\pm$0.1 & 1.2$\pm$0.1\\
  GJ\,1151 & M4.5 & 8.04 & $0.52\pm$0.06 & 64$\pm$6 &  1,\,4 & 125$\pm$23 & 1.4$\pm$0.2 \\
  GJ\,450 & M1.5 & 8.76 & $0.37\pm$0.05 & 62$\pm$8 & 1,\,2 & 23$\pm$1 & 2.0$\pm$0.3 \\
  LP\,169-22 & M5.5 & 10.49 & $-0.48\pm$0.07 & 62$\pm$11 & 1,\,3&  & 4.4$\pm$0.9\\
  CW\,UMa & M3.5 & 13.36 & $-1.21\pm$0.11 & 61$\pm$4 & 1,4 & 7.77 & 3.4$\pm$0.3 \\
  HAT\,182-00605 & M4.0 & 17.87 & $-0.59\pm$0.08 & 76$\pm$7 & 1,\,3  & 2.21 & 1.4$\pm$0.2\\
  LP\,212-62 & M5.0 & 18.20 & $-6.43\pm$0.17 & 90$\pm$6 & 4,\,4 & 60.75 & 69.7$\pm$0.1\\
  DG\,CVn$^{*}$ & M4.0 & 18.29 & $-0.57\pm$0.06& 91$\pm$9 & 2,\,2  & 0.11 &  1.0$\pm$0.6 \\ 
  GJ\,3861 & M2.5 & 18.47 & $0.62\pm$0.04 & 70$\pm$6 &  2,\,2 &  & 1.6$\pm$0.2\\
  CR\,Dra$^{*}$ & M1.5 & 20.26 & $8.12\pm$0.14 & 92$\pm$8 & 20,\,21 & 1.98 & 5.2$\pm$0.6 \\ 
  GJ\,3729$^{*}$ & M3.5 & 23.57 & $-1.04\pm$0.09 & 73$\pm$3 &  1,\,3 & 13.59 & 3.4$\pm$0.5\\ 
  G\,240-45 & M4.0 & 27.59 & $1.08\pm$0.08  & 80$\pm$4 & 1,\,3 &  & 40$\pm$5\\
  2MASS\,J09481615+5114518 & M4.5 &36.17 & $1.76\pm$0.09 &  96$\pm$4 & 1,\,3 &  & 92$\pm$9\\ 
  LP\,259-39 & M5.0 & 36.93 & $-0.41 \pm 0.09$ & 67$\pm$6 & 1,1 & & 20$\pm$4\\ 
  2MASS\,J10534129+5253040 & M4.0&45.19 & $-0.70\pm$0.10 &  60$\pm$7 & 1,\,3 &  & 15$\pm$3\\ 
  2MASS\,J14333139+3417472 & M5.0&47.84 & $0.68\pm$0.11 &  61$\pm$5 & 1,\,1 &  & 140$\pm$30\\  
    \hline\end{tabular}
\end{center}
\end{table}

All of our sources, except AD\,Leo and DO\,Cep, show a high degree of circular polarisation ($>$\,60\%), high brightness temperature ($T_{B} \gtrsim 10^{12}$\,K), broadband emission ($\Delta \nu / \nu \sim 1$, where $\nu$ is frequency), and are observed to be emitting for $\gtrsim$4 hours in multiple fields. We demonstrate (see Supplementary Information section ``Radio emission mechanisms'') that depending on the coronal and chromospheric activity of the star, plasma or ECMI mechanisms are the only processes that can generate the long-lived low-frequency emission with such high brightness temperature and circularly polarised fraction\cite{harish2019}. The detected emission properties of our sample show a clear departure from the characteristics of gigahertz-frequency emission from most main-sequence stars, which is often either hours-long with low or zero polarisation, or consists of short-duration ($\lesssim 30$\,min), highly-polarised bursts\cite{2019ApJ...871..214V,2008ApJ...684..644H,Zic2019}.

The low-frequency emission from stars with low X-ray luminosities and long ($\gtrsim 2$\,d) rotation periods is more likely to be generated by ECMI, resembling the emission processes seen in the magnetospheres of sub-stellar objects such as ultracool dwarfs and gas-giant planets\cite{2004JGRA..109.9S15Z,1998JGR...10320159Z,2008ApJ...684..644H}. For the stars in our sample that are coronally active and rapidly rotating (e.g. AD\,Leo, DO\,Cep), plasma emission could plausibly generate the observed radio emission.

With the first blindly formed sample of coherently-emitting low-frequency stellar systems, we can investigate the potential origins of the emission by identifying any trends with stellar properties. In particular, it is unclear how the coherent emission is related to the coronae of the stars. Radio emission associated with known radio stars is powered by high-energy supra-thermal electrons with energies of $\sim 10\,{\rm keV} - 1\,{\rm MeV}$\cite{1985ARA&A..23..169D,Gudel2002}. In stellar coronae, the electrons are accelerated by the chromospheric and coronal dissipation of magnetic fields\cite{1985ARA&A..23..169D,Gudel2002}. This is evidenced by a tendency of chromospherically-active stars to be radio detected, a fact that has been used to probe the stellar dynamo\cite{Gudel2002}. In sub-stellar objects that lack a dominant corona, like brown dwarfs and gas giants, the electrons are thought to be accelerated at large distances from the central object due to either a breakdown of co-rotation between the object's magnetic field and its magnetospheric plasma, or an interaction with an orbiting satellite\cite{2018A&A...618A..84Z,2008ApJ...684..644H}. 

\begin{figure}
   \centerline{\resizebox{\textwidth}{!}{\includegraphics{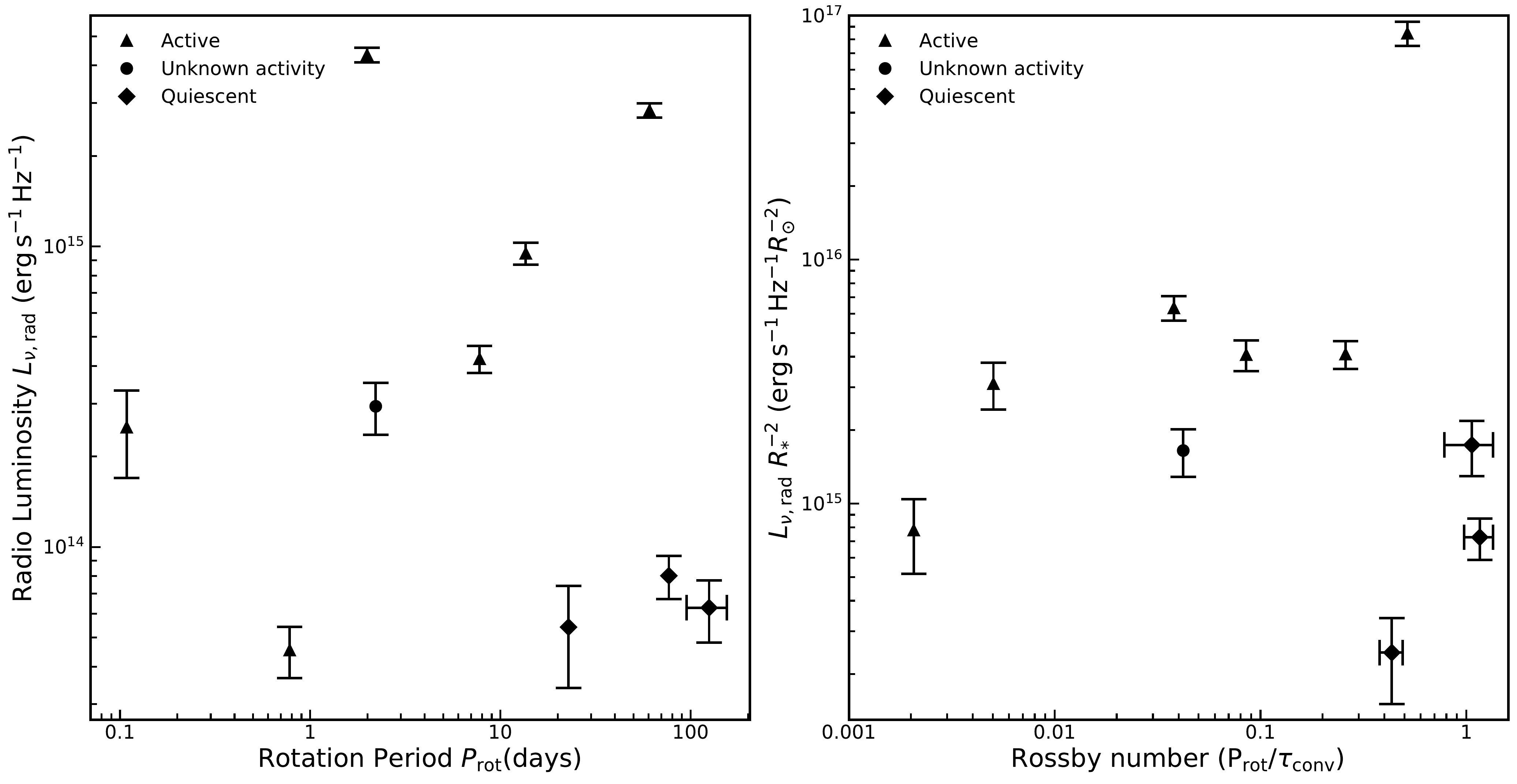}}}
 \caption{\label{fig:period_radlum}
 {\bf 144\,MHz radio luminosity $L_{\nu,\mathrm{rad}}$ as a function of rotation period $P_{\mathrm{rot}}$ (left panel), and radio surface flux density ($L_{\nu,\mathrm{rad}} / R^{2}_{*}$, where $R_{*}$ is stellar radius) as a function of Rossby number $R_{\mathrm{o}}$ (right panel), for ten of our M~dwarf stellar systems.}
 Chromospherically active, quiescent, and unknown chromospheric activity stellar systems are plotted as triangles, diamonds, and circles, respectively. We find no correlation between $L_{\nu,\mathrm{rad}}$ and $P_{\mathrm{rot}}$, or with $L_{\nu,\mathrm{rad}}/R^{2}_{*}$ and $R_{\mathrm{o}}$, with 95\% confidence. The uncertainties in $P_{\mathrm{rot}}$ and $R_{\mathrm{o}}$ are smaller than the symbol size unless otherwise indicated. The error bars represent 1-$\sigma$ uncertainty.} 
\end{figure}

To determine whether the coherent radio emission we detect is powered by processes in the corona or chromosphere, we compare the radio luminosity of our sample with known coronal and chromospheric activity indicators that have been used to understand the origin of gigahertz-frequency stellar emission. We find a conspicuous lack of correlation between the radio luminosity of our sample with well-known chromospheric activity indicators such as H$\alpha$ luminosity, rotation period, and Rossby number as shown in Figure\,\ref{fig:period_radlum} and the Supplementary Information. The Kendall-$\tau$ rank correlation test rejects a relationship between the radio luminosity and rotation period, as well as between radio surface flux density and the Rossby number, at $\approx 95\%$ significance. 

The lack of correlation with the Rossby number is interesting because previous gigahertz-frequency observations have found a single trend of radio surface flux density with Rossby number that ranges from Sun-like G2 stars to L4 brown dwarfs\cite{2012ApJ...746...23M,1989MNRAS.236..129S,1988AJ.....96..371S}. Our low-frequency detected M~dwarf sample spans over three orders of magnitude of Rossby number, with the largest Rossby numbers ever observed for radio-bright M~dwarfs. Only substantially larger and slower-rotating G~dwarfs have been detected in the radio with such large Rossby numbers\cite{2012ApJ...746...23M}. We also do not predominantly detect rapid rotators, unlike gigahertz-frequency observations, with a constant detection rate for stars with rotation periods from 0.1 to 100 days (see Supplementary Information Section ``Rotation and detection rate''). This is in contrast to the canonical expectation that radio bright stars are chromospherically active and rapidly rotating\cite{2012ApJ...746...23M,1988AJ.....96..371S}. While Figure\,\ref{fig:period_radlum} likely has stars producing low-frequency emission via plasma and ECMI processes, the combination of the lack of correlation of radio surface flux density with Rossby number, the largest Rossby numbers for radio-bright M\,dwarfs, and the independence of our detection rate with stellar rotation period supports the idea that the low-frequency radio emission we detect is different than that commonly detected for M and ultracool dwarfs at gigahertz frequencies\cite{2012ApJ...746...23M}.Additionally, our sample also shows no clear dependence on the coronal temperature and density traced by soft X-ray luminosity, as has also been shown for ultracool dwarfs\cite{2014ApJ...785....9W} (see Supplementary Information Section ``G{\"u}del-Benz relationship''). 

It is also important to note that a high coronal temperature does not immediately imply that plasma emission is producing the low-frequency emission. For example, all 20 detections of CR\,Dra have an identical circularly polarised handedness, implying that the emission site is situated in a stable magnetic field arrangement\cite{callinghamCRDra}. Plasma emission would be expected to flip in circularly polarised handedness as it emerges from areas of complex magnetic geometry\cite{2019ApJ...871..214V}. Therefore, even though CR Dra does have a large coronal base density, we suggest that we are observing ECMI from this star. In fact, for all of the stars of which we have multiple detections (Table\,\ref{tab:radio_detections}), no star has flipped circular polarity. Consistent circularly-polarised handedness of persistent radio emission has also previously been observed on some active stars at gigahertz-frequencies, and has been used to support ECMI as the emission mechanism\cite{2019ApJ...871..214V}.

The lack of a radio luminosity -- stellar activity relationship, and the likelihood that the low-frequency emission is generated by ECMI in our quiescent stars, suggests that the radiation from the quiescent members of the population has similarities to auroral radio emission seen in sub-stellar objects. In particular, a coherent emission mechanism appears ubiquitous across the M~dwarf main sequence and independent of whether the core of a M~dwarf is partially or fully convective. Taken together, these two facts suggest that the electron acceleration site for some of our stars could be extrinsically sourced and that the emission could be a stellar analogue of the radio aurorae seen in brown dwarfs and gas-giant planets. 

This sub-stellar auroral behaviour in M~dwarfs is not unexpected. M~dwarfs are known to retain strong magnetic fields over gigayear timescales\cite{2010MNRAS.407.2269M}. As a result, their Alfv\'{e}n surfaces likely extend to tens of stellar radii. Unlike the Sun, whose ECMI radiation is powered by processes within the corona, auroral processes in M~dwarfs can transport energy from within the Alfv\'{e}n surface back to the star to power intense ECMI emission. We note that one of our detections, WX\,UMa, provides direct evidence for a distant source site. WX\,UMa has a large-scale dipolar magnetic field with a strength of $\sim 3.5\,{\rm kG}$ at the stellar surface\cite{2010MNRAS.407.2269M} (cyclotron frequency of $\approx 10\,{\rm GHz}$) which means that the 120\,MHz emission from WX\,UMa must originate at a distance of at least 5.5 stellar radii from the star\cite{2021arXiv210501021D}. Such a conclusion has the important implication that M~dwarf coronae can be structurally similar to planetary magnetosphere in that they are dominated by a global dipole with a closed field topology.

Our sample likely consists of systems that are producing low-frequency emission via either plasma or ECMI processes. For the systems emitting via ECMI, there are two known mechanisms that power hours-long radio emission: co-rotation breakdown and interaction with a conducting and/or magnetised obstacle\cite{2008ApJ...684..644H,2012ApJ...760...59N}. ECMI emission from co-rotation breakdown is powered by stellar rotation and has been used to explain the auroral emission from Jupiter to brown dwarfs and some ultra-cool dwarfs (M7 and later type)\cite{2008ApJ...684..644H}. If some of the stars in our sample have magnetospheric parameters comparable to ultra-cool dwarfs, a simple empirical scaling of the radiated power with rotation period shows that this mechanism is only feasible for a small sub-sample of our stars which have rapid rotation (period $\lesssim 2\,{\rm day}$)\cite{2012ApJ...760...59N} (see also Supplementary Information Sections ``Radio emission mechanisms'' and ``Rotation period and detection rate''). 

Therefore, the ECMI mechanism for stars in our sample with long rotation periods could be driven by star-exoplanet  interactions\cite{2004ApJ...612..511L,2007P&SS...55..598Z} or by a hitherto unknown energy release mechanism, possibly from sites in the equatorial current sheets that extend into interplanetary space\cite{1992ApJ...393..341L}. Exoplanet-driven ECMI has been theoretically modelled based on our understanding of Jupiter-Io interactions\cite{2007P&SS...55..598Z,harish2019}. 
The radio flux density of our sample is consistent with theoretical expectations for interaction with the most common type of terrestrial planets found around M~dwarfs\cite{2018ApJ...854...72T} (see Supplementary Information Section ``Radio emission mechanisms''). Although such planets are ubiquitous around M~dwarfs, the small detected fraction could be explained by the angular beaming of ECMI emission combined with the requirement that only M~dwarfs with surface magnetic field strengths larger than about $50$\,Gauss can generate emission at our observation frequency\cite{2019MNRAS.488..633V}(see Supplementary Information Section ``Variability''). However, only follow-up studies to test if periodicity is evident in the radio emission from these stars will be able to validate such a suggestion.

Our sample can be interpreted along with previous gigahertz-frequency studies\cite{2019ApJ...871..214V,2008ApJ...684..644H}, to arrive at a unified phenomenology of radio emission from M~dwarfs. We suggest that M~dwarfs, unlike the Sun, can harbour both star-like and gas-giant-like behaviour in their radio emission. Their coronae likely consist of short X-ray bright coronal loops that flare occasionally, and a more tenuous and extended coronal gas supported by a large-scale magnetic field. The polarized short-duration bursts commonly detected at gigahertz frequencies likely originate in flaring coronal loops and therefore trace coronal and chromospheric flaring activity seen at other wavelengths\cite{Gudel2002}. On the other hand, the long-duration polarised emission we are detecting originates in the large-scale corona (outside the loops), from a quasi-continuous acceleration mechanism  similar to that seen on Jupiter.

This unified model does not restrict the presence of ECMI emission to low frequencies. In highly-active M~dwarfs with strong large-scale magnetic fields, the emission can extend to the gigahertz regime. Such emission has been recently detected in dMe-type flare stars with kilogauss-level magnetic fields\cite{2019ApJ...871..214V,Zic2019}. Most previous efforts were restricted to highly active stars in which the distinction between continuous acceleration due to frequent coronal flaring and an extra-coronal process as proposed here is difficult to discern. The untargeted character of our detection strategy has allowed us to infer that such coherent emission processes are not isolated to chromospherically-active, rapidly-rotating M~dwarfs.

Our model can be tested via long-term monitoring of the sample in different spectral bands. Co-rotation breakdown in our fast-rotating stars will generate emission that is modulated at the rotation period of the star. On the other hand, exoplanet-induced emission must show similar modulation but over the orbital period of the putative planet, which can be confirmed by radial-velocity observations. In such a situation, it will be possible to assess the magnetic field strength of both the star and exoplanet, the latter being possible if the large-scale magnetic topology of the star is known and if the star-planet interaction is dipolar\cite{2007P&SS...55..598Z,Strugarek2016}. Therefore, some of our stars in this sample represent excellent candidates for searching for star-exoplanet interaction signatures.

\bibliographystyle{naturemag}

\newpage

\section*{Acknowledgements}
The LOFAR data in this manuscript were (partly) processed by the LOFAR Two-Metre Sky Survey (LoTSS) team. This team made use of the LOFAR direction independent calibration pipeline (\url{https://github.com/lofar-astron/prefactor}), which was deployed by the LOFAR e-infragroup on the Dutch National Grid infrastructure with support of the SURF Co-operative through grants e-infra 170194 e-infra 180169\cite{2017isgc.confE...2M}. The LoTSS direction dependent calibration and imaging pipeline (\url{http://github.com/mhardcastle/ddf-pipeline/}) was run on compute clusters at Leiden Observatory and the University of Hertfordshire, which are supported by a European Research Council Advanced Grant [NEWCLUSTERS-321271] and the UK Science and Technology Funding Council [ST/P000096/1]. The J\"ulich LOFAR Long Term Archive and the German LOFAR network are both coordinated and operated by the J\"ulich Supercomputing Centre (JSC), and computing resources on the supercomputer JUWELS at JSC were provided by the Gauss Centre for Supercomputing e.V. (grant CHTB00) through the John von Neumann Institute for Computing (NIC). This work was performed in part under contract with the Jet Propulsion Laboratory (JPL) funded by NASA through the Sagan Fellowship Program executed by the NASA Exoplanet Science Institute. JRC thanks the Nederlandse Organisatie voor Wetenschappelijk Onderzoek (NWO) for support via the Talent Programme Veni grant. PNB and JS are grateful for support from the UK STFC via grant ST/R000972/1. RJvW acknowledges support from the Vidi research programme with project number 639.042.729, which is financed by the Netherlands Organisation for Scientific Research (NWO). AD acknowledges support by the BMBF Verbundforschung under the grant 05A17STA. This research has made use of the SIMBAD database, operated at CDS, Strasbourg, France, and NASA's Astrophysics Data System. This work has also made use of TOPCAT\cite{2005ASPC..347...29T}, the \textsc{IPython} package\cite{PER-GRA:2007}; SciPy\cite{scipy};  \textsc{matplotlib}, a \textsc{Python} library for publication quality graphics\cite{Hunter:2007}; \textsc{Astropy}, a community-developed core \textsc{Python} package for astronomy\cite{2013A&A...558A..33A}; and \textsc{NumPy}\cite{van2011numpy}. 

\textbf{Author Contributions:} JRC initiated the LOFAR project that led to the discovery of the sources, conducted the crossmatching analysis, processed the extracted source visibilities, and wrote the manuscript. JRC and HKV developed the detection strategy. HKV led the theoretical interpretation of the detections and contributed substantially to the manuscript. TWS processed the radio data with LoTSS DR2 data reduction software developed by TWS, MJH, CT, WLW, FdG, JRC and other members of the LoTSS survey collaboration. BJSP helped develop the project and contributed to the manuscript. IED helped process individual stellar system datasets. JS and PNB processed the ELAIS-N1 deep-field data. HJAR is the principal investigator of the broader LOFAR Two-Metre Sky Survey. RJvW developed the extraction and re-calibration pipelines which optimized the calibration towards a target of interest. PZ provided comprehensive feedback on the manuscript. All authors commented on the manuscript.

\textbf{Competing interests:} The authors have no competing interests with respect to this manuscript.

\textbf{Data availability:} LOFAR visibilities taken before 2020 are publicly available via the LOFAR Long Term Archive. The \emph{XMM-Newton} data on LP\,212-62, G\,240-45, LP\,169-22, and GJ\,1151 are available through the \emph{XMM-Newton} Science Archive (XSA). All other data used in the manuscript have been sourced from the public domain. 

\textbf{Code availability:} The important codes used to analyse the data are available at the following sites:
    WSClean (\url{https://gitlab.com/aroffringa/wsclean})
    and DDF pipeline (\url{https://github.com/mhardcastle/ddf-pipeline}).
    
\textbf{Correspondence:} Correspondence and requests for materials should be addressed to JRC (email: jcal@strw.leidenuniv.nl).

\newpage

\section{Methods}
\setcounter{figure}{0}
\subsection{LoTSS circular polarisation data processing and source detection}

Moderate resolution (20$''$), wide-band (120 to 168\,MHz) circular polarisation dirty images are standard products from the LoTSS-DR2 pipeline (described briefly in Sec.\ 5.1 of Shimwell et al.\cite{2019A&A...622A...1S} and by Tasse et al.\cite{Tasse2021}) which, to date, have been used to image approximately 1000 LoTSS pointings covering $\approx$20\% of the Northern sky. The 8\,hour LoTSS pointings are observed with 1\,s integration and 12.5\,kHz spectral resolution. The flux density scale of the images is corrected by matching sources in each field with NVSS and then assuming an average spectral index of sources. The median sensitivity of the Stokes~V images is 80$\mu$Jy\,beam$^{-1}$. 

Each individual 20$''$ Stokes~V image was searched for $\geq$4$\sigma$ sources using the Aegean source finder\cite{2012MNRAS.422.1812H,2018PASA...35...11H}, where $\sigma$ is local-rms noise. To ensure a detection was reliable and not the product of a noise spike, any significant Stokes~V source was required to be co-incident with a Stokes~I source in a 6$''$ image, as the corresponding Stokes~I image has independent noise properties.

To determine the amount of Stokes~I to V leakage, we measured the flux density at the brightest Stokes~V pixel co-incident with the position of all bright ($>$\,125\,mJy), compact ($<6''$ largest angular size) Stokes~I sources. The corresponding Rician distribution peaks at a Stokes~I to V ratio of $|V/I| = (0.14 \pm 0.07)\%$. Based on this, we consider any $\geq4\sigma$ Stokes~V source with $|V/I| >1\%$ as a reliable Stokes~V detection. While we are confident that we have accurately characterised the leakage into Stokes~V, to be conservative we exclude 2$'$ regions around sources with a Stokes~I flux density $>$125\,mJy. Such a cut does not mean we are missing the bright end of the stellar system population, as evidenced by the resulting source counts distribution (see Supplementary Information Section ``144\,MHz source counts of M dwarf stellar systems'').

Once significant and reliable Stokes~V emission was identified, a post processing step was carried out to enable us to study the emission in more detail. After running the LoTSS-DR2 pipeline, all sources from a given LoTSS pointing were subtracted with the direction dependent calibration solutions, apart from sources in small region around the target of interest. The datasets were then phase shifted towards the region of interest and corrected for the LOFAR station beam attenuation. These steps are followed by three cycles of ``TEC and phase'' self-calibration on 10-20~sec timescales, which helped correct for any ionospheric distortions by solving for variations in total electron content (TEC). That step was followed by seven rounds of diagonal gain calibration on timescales of 10-60~min, depending on the amount of source flux density available. The ``TEC and phase'' solutions were pre-applied before solving for the diagonal gains. DPPP was employed for the calibration\cite{2018ascl.soft04003V} and the images were made using WSClean\cite{2014MNRAS.444..606O}. The final products are small ($<$5\,GB) datasets for each pointing that are calibrated in the direction of the target and with sources away from the target region subtracted. 

All images were made with a robust parameter of $-0.5$ and the bandwidth/time-resolution was set to ensure that the source was $\gtrsim3\sigma$ in each image. The Stokes~I and V images for all our detections are available in Supplementary Information Figure\,\ref{fig:rouges}.

\begin{figure*}
\renewcommand\figurename{Supplementary Information Figure}
   \centerline{\scalebox{0.32}{{\includegraphics{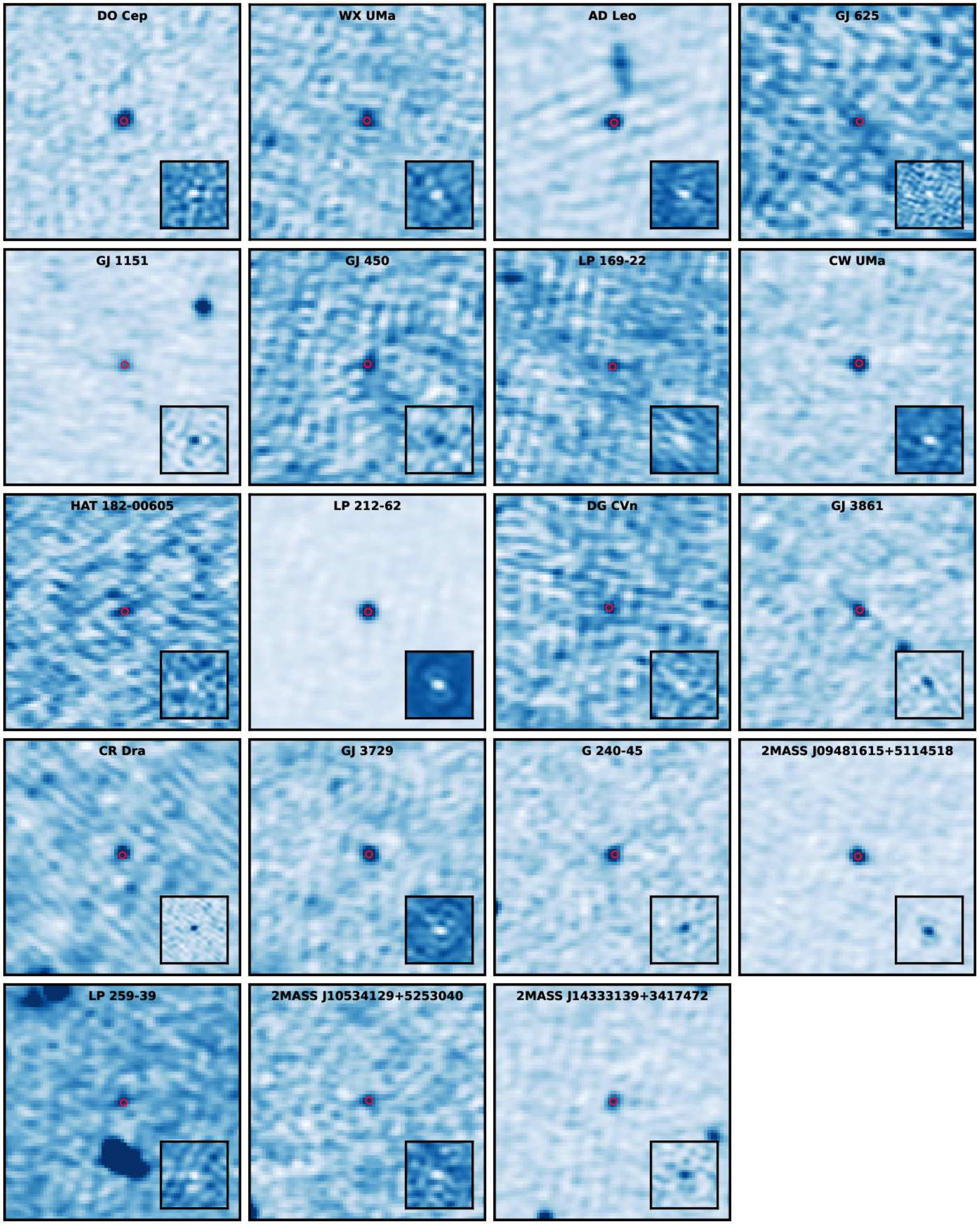}}}}
 \caption{\label{fig:rouges} 
 {\bf Total intensity (Stokes I) and circularly polarised (Stokes~V) (insets) 144\,MHz images of the detected M~dwarf systems.} 
The red circle in each image represents the position of the \emph{Gaia} DR2 counterpart at the time of the LOFAR observation. The radius of the circle is 1$''$. The intensity scale of each image is different, with dark-blue representing positive flux density. All detections are $\geq 4\sigma$ in both Stokes~I and V. The Stokes~I images are deconvolved and have a 5$''$ point-spread function. The Stokes~V images are not deconvolved and have a 20$''$ diameter point-spread function. The horizontal and vertical axes of each image are right-ascension and declination, respectively. } 

\end{figure*}

All LoTSS fields neighbouring to a confirmed detection were also inspected to test if the source was additionally detected in a different epoch. The source was considered undetected in a neighbouring field if the local rms was comparable to or better than the local rms of the field in which the source was detected, as expressed in Table\,\ref{tab:radio_detections}. On average, this involved inspecting 2 to 3 fields per detection. CR\,Dra and GJ\,625 are part of the ELAIS-N1 deep field which has 21 independent 8\,hour LOFAR observations taken at approximately the same local sidereal time.

\subsection{ROSAT and other X-ray flux measurements}

The majority of our low-frequency detections have a counterpart in the Second ROSAT all-sky survey (2RXS) source catalogue\cite{2016A&A...588A.103B}. We converted the ROSAT count rate to a 0.1 to 2.4\,keV flux using the conversion factor\cite{1995ApJ...450..401F,1995ApJ...450..392S} $CF = (5.30\mathrm{HR1} + 8.31) \times 10^{-12}$ erg\,cm$^{-2}$\,counts$^{-1}$, where $\mathrm{HR1}$ is the hardness ratio in the soft ROSAT band.

For the sources that did not have a counterpart in 2RXS, we searched if they had been observed by \emph{XMM-Newton} or \emph{Chandra}. Any flux measurements or upper limits were converted to the 0.1 to 2.4\,keV energy band. 2MASS\,J14333139+3417472 was detected by \emph{Chandra} several times in a survey of the Bootes field. We report the flux listed in the Chandra Source Catalog\cite{2010ApJS..189...37E}. 

\emph{XMM-Newton} observed LP\,212-62 serendipitously for $\approx$\,30\,ks at $\approx 2.5'$ away from the pointing centre on 2007-11-11 (ObsID: 0503630201). We reduced the observation following standard routines\cite{Callingham2012,2019NatAs...3...82C}. The  observation data were ingested using the \emph{XMM-Newton} Science
Analysis System (version 16.0.0), and background flares were filtered from the European Photon Imaging Camera data. A circular source region of radius $\approx16''$ was used for spectral and timing extraction, with background spectra and timing extracted on the same CCD from a region with twice the area. Photon redistribution matrices and ancillary response files were constructed for the spectra and lightcurve for the pn, MOS1, and MOS2 cameras using standard calibration files.

The average 0.2 to 10.0\,keV count rate for LP\,212-62 during the observation was $\approx 0.08$ count\,s$^{-1}$, but for $\approx$30\,min the source flared to a count rate of $\approx 0.25$ count\,s$^{-1}$. This flare was filtered before constructing the quiescent spectrum of LP\,212-62. The quiescent spectrum of LP\,212-62 was best fit by a single photoelectrically absorbed optically-thin plasma model (XSPEC\cite{Arnaud1996} model $\tt{apec}$) with a temperature of $0.41 \pm 0.04$\,keV. Such a model provides a 0.2 to 2.4\,keV flux of $(9.5\pm 0.7) \times$10$^{-14}$\,erg\,s$^{-1}$\,cm$^{-2}$, consistent with the 2RXS upper-limit at the position of LP\,212-62.

GJ\,1151 was also observed by \emph{XMM-Newton} for $\approx$\,10\,ks on 2018-11-01 (ObsID: 0820911301). The data reduction performed was identical to the process outlined for LP\,212-62, with the exception that we discarded the pn data as GJ\,1151's position on a chip edge meant the total effective area is not well calibrated by the standard pipeline. The average 0.2 to 10.0\,keV count rate for GJ\,1151 was $\approx 0.015$ count\,s$^{-1}$, and we did not detect any X-ray flares. The spectrum was reasonably fit by an $\tt{apec}$ model with a temperature of $0.25^{+0.09}_{-0.06}$\,keV. The estimated 0.2 to 2.4\,kev flux from such a model is (2.0$\pm$1.1)\,$\times 10^{-14}$\,erg\,s$^{-1}$\,cm$^{-2}$, consistent with the non-detection of GJ\,1151 in an earlier 3\,ks \emph{Chandra} pointing\cite{2018MNRAS.479.2351W}.

Finally, we observed G\,240-45, LP\,169-22, and 2MASS\,J09481615+5114518 with \emph{XMM-Newton} for $\approx$\,15, 13, and 8\,ks, respectively (ObsIDs: 0864480301, 086448101, 0864480201; PI: Callingham). Following identical data reduction procedures to those described above, we detect G\,240-45 at $\approx$4\,$\sigma$ in the pn data. It is not detected in the individual MOS cameras. We do not detect any flares in the pn data. The spectrum is best fit by an $\tt{apec}$ model with a temperature of $0.3^{+0.3}_{-0.1}$\,keV. The estimated 0.2 to 2.4\,kev flux from such a model is (1.1$\pm$0.9)\,$\times 10^{-15}$\,erg\,s$^{-1}$\,cm$^{-2}$. Therefore, even though G\,240-45 is $\approx$3.5 times further away than GJ\,1151, it has a similar X-ray luminosity. 

2MASS\,J09481615+5114518 was detected in both pn and MOS data. There was no clear flares evident in the pn lightcurve. The spectrum is best fit by an $\tt{apec}$ model with a temperature of $0.14^{+0.04}_{-0.05}$\,keV. The estimated 0.2 to 2.4\,kev flux from such a model is (2.1$\pm$0.8)\,$\times 10^{-14}$\,erg\,s$^{-1}$\,cm$^{-2}$.

LP\,169-22 was not detected in our the XMM-Newton observation. While close to the X-ray detected star 
TYC\,3451-1027-1, we derive a conservative three-sigma 0.2 to 2.4\,kev flux upper-limit of 1.8\,$\times 10^{-14}$\,erg\,s$^{-1}$\,cm$^{-2}$. This corresponds to an X-ray luminosity $<$2.3\,$\times 10^{26}$\,erg\,s$^{-1}$, making LP\,169-22 a quiescent candidate of our sample.

\subsection{Radio source association with \emph{Gaia} DR2}

A benefit of wide-field radio surveys is that implicit biases inherent to targeted observations are bypassed, ensuring we observe a more complete sample of the stellar population outside of active stars and close binaries\cite{Gudel2002}. However, even with the LoTSS astrometric precision of $\approx$\,0.2$''$ to 0.5$''$, a crossmatch of total-intensity (Stokes~I) and Galactic \emph{Gaia} Data Release 2 (DR2) sources is dominated by chance co-incidence matches\cite{2019RNAAS...3...37C}.

To form a reliable sample of Galactic \emph{Gaia}-LoTSS counterparts, we restricted the crossmatching sample to LoTSS sources that were $\geqslant4\sigma$ in Stokes~V and $\geqslant4\sigma$ in Stokes~I, where $\sigma$ is the local rms noise. Restricting our search to only circularly-polarised sources also suited our science goal of identifying coherent low-frequency-emitting stellar systems. The source density of the Stokes~V radio sky is orders of magnitudes lower than the Stokes~I radio sky because $>99\%$ of extragalactic sources are not circularly polarised. Only stellar systems, ultracool dwarfs, and pulsars are known to have circular-polarisation fractions $>10\%$\cite{1998MNRAS.301..235G,1998MNRAS.300..373H,2019ApJ...871..214V}. We find $\approx0.1$ circularly-polarised source per sq.~degree with a circularly-polarised fraction $>2\%$, compared to $\approx770$ Stokes~I sources per sq.~deg. 

After identifying circularly-polarised LoTSS sources, we cross-matched their Stokes~I position to all proper-motion corrected \emph{Gaia} DR2 sources within 1$''$ that had a parallax over error of $\varpi / \varpi_{\mathrm{err}} \geqslant 3$. Since all the LoTSS pointings searched are at Galactic latitudes $|b| > 20^{\circ}$, the average \emph{Gaia} DR2 source density for sources with $\varpi / \varpi_{\mathrm{err}} \geqslant 3$ is $\approx$\,1300 per sq.~degree. Therefore, we would expect the number of chance-coincidence associations between a Gaia Galactic source and the Stokes\,V LoTSS sample to be $\approx0.05$. As performed by Callingham et al.\cite{2019RNAAS...3...37C}, the chance-coincidence rate was also empirically confirmed by applying 10$'$-20$'$ shifts to the \emph{Gaia} DR2 positions and performing the crossmatch again. All of the sources presented in this paper were also observed to display variability incongruent with noise or proper motion shifts in neighbouring LoTSS fields. Therefore, all detections are considered reliable. 

The above \emph{Gaia} DR2 $\varpi / \varpi_{\mathrm{err}} \geqslant 3$ requirement was to ensure we were selecting Galactic sources and to remain as unbiased as possible to the type of stellar systems detected. We did not restrict ourselves to crossmatching to only known M\,dwarfs so that we could systematically explore the type of sources that occupy the highly-circularly polarised Galactic radio sky. Hence, we have also detected other types of stellar systems such as RS Canum Venaticorum (RS\,CVn), BY Draconis (BY\,Dra), FK Comae Berenices (FK\,Com), and W Ursae Majoris (W\,Uma) variable stars, millisecond pulsars, and others. For this publication we have chosen to focus on the population of non-degenerate, main-sequence stellar systems that do not have known binary companions with $<1$\,AU separation. We have not detected any isolated, main-sequence star other than M~dwarfs. For the area of sky we have observed, the non-detection of any isolated, main-sequence star is consistent with the number density of M~dwarfs relative to G to K\,dwarfs stars within 50\,pc\cite{Henry2006}. We also note that we have not detected any previously identified radio-bright ultracool dwarfs, despite there being more than 20 candidates in our survey footprint\cite{2012ApJ...746...23M}. This is likely because the majority of the searched ultracool dwarfs only have observed flux densities $<300$\,$\mu$Jy at gigahertz frequencies. We are following up highly circularly polarised detections that have no optical counterparts. 

\subsection{\emph{Gaia} DR2 M~dwarf comparison sample}

To derive the detection statistics for the M~dwarf population presented Figure\,\ref{fig:ms_dectstats}, we have to isolate a general representative population. We take our sensitivity horizon to be 50\,pc, our most distant detection. After selecting all \emph{Gaia} DR2 sources within 50\,pc that are in the area so far surveyed by LoTSS, we followed the quality cuts outlined by Kiman et al.\cite{2019AJ....157..231K} to form a clean sample of M~dwarfs contained in \emph{Gaia} DR2.

Firstly, we ensured that our sample contained stars with accurate astrometric data by requiring $\varpi / \varpi_{\mathrm{err}} > 10$ and that the number of visibility periods used were $> 8$. Secondly, to reliably select red stars we made a photometric quality cut in the red part of the \emph{Gaia} band $G_{\mathrm{RP}}$ ([630,1050]\,nm) by requiring a signal-to-noise ratio of $>10$. This resulted in a final sample of 4518 stars with \emph{Gaia} colour $G - G_{\mathrm{RP}} > 0.9$. All of our radio detected stars are contained in this final general sample. 

Note that we did not make a quality cut on the \emph{Gaia} ``unit weight error'' (UWE) as we wanted to remain sensitive to detecting unresolved binaries. The majority of the scatter around the main-sequence in the top panel of Figure\,\ref{fig:ms_dectstats} is because we have excluded this UWE quality cut.

\subsection{Rossby number calculation}

The Rossby number is a common parameter used to compare coronal and chromospheric activity strengths across both mass and rotation period ranges. To calculate the Rossby number for our detections, we require the rotation period of the star $P_{\mathrm{rot}}$ and the convective turnover time $\tau_{\mathrm{conv}}$. The stellar rotation period can be measured directly from photometric observations\cite{2017ApJ...834...85N,2019A&A...621A.126D}. To estimate the convective turnover time, we used the empirical values derived using X-ray luminosity and $V - K_{s}$ colour for partially and fully convective active M~dwarfs by Wright et al.\cite{2018MNRAS.479.2351W}.

\subsection{Circular polarisation sign convention}

There are several instances in the collection, correlation, and reduction of LOFAR data where a sign-convention decision can influence the final handedness of our Stokes~V measurements. To add to further to the confusion, there are many competing definitions used in astronomy\cite{2018arXiv180607391R,2010PASA...27..104V}. Pulsar astronomy commonly defines the sign of circularly-polarised light as left-hand circularly-polarised light minus right-hand circularly-polarised light. To ensure that the handedness of our Stokes~V measurements are consistent with that definition\cite{2010PASA...27..104V}, we checked whether the Stokes~V signs of pulsars observed in LoTSS were consistent with their circular polarisation handedness reported in the literature\cite{1998MNRAS.301..235G,1998MNRAS.300..373H,2010PASA...27..104V}. While we are integrating over many pulse profiles and comparing the pulsar's polarisation properties at a lower frequency than originally measured, pulsars are one of the few sources outside of stellar systems that can be strongly circularly polarised at low-frequencies. 

In the LoTSS data we have reduced, we have detected significant circular polarisation from nine pulsars that are also in Han et al.\cite{1998MNRAS.300..373H}, Gould et al.\cite{1998MNRAS.301..235G}, and van Straten et al.\cite{2010PASA...27..104V}: B0917+63, B0025+33, B2210+29, B0751+32, B2315+21, B0655+64, B0823+26, B1737+13, and B0301+19. We exclude B0751+32 and B2315+21 from our analysis as their circularly-polarised handedness reverses between two of the references\cite{1998MNRAS.301..235G,1998MNRAS.300..373H}. For the remaining seven pulsars, we are in agreement with the handedness reported in the literature, implying that our Stokes~V sign convention is consistent with the definition applied in pulsar astronomy.

\section{Supplementary Information}

\subsection{Literature information on detected stellar systems:}

The multi-wavelength properties pertinent to this manuscript for our sample are presented in Supplementary Table\,\ref{tab:literature_info}. 

\newpage

\setcounter{table}{0}
\begin{table*}
\renewcommand\tablename{Supplementary Information Table}
\small
  \caption{\label{tab:literature_info} {\bf Multi-wavelength properties of the 144-MHz detected M~dwarfs in increasing distance.} $M_{*}$, $R_{*}$, $L_{\nu,\mathrm{rad}}$, $L_{X}$, $L_{\mathrm{H}\alpha}$/$L_{\mathrm{bol}}$, $T_{B,\mathrm{plasma}}$ and Epoch correspond to stellar mass, stellar radius, 144\,MHz radio luminosity, 0.2-2\,keV X-ray luminosity, ratio of H$\alpha$ to bolometric luminosity, the brightness temperature of fundamental plasma emission when a hot component 30 times the relative coronal temperature is injected, and the epoch of radio detection, respectively. If multiple detections are known, the earliest detection is listed. Stars with known close ($<$\,10\,day orbit) exoplanets are identified by an asterisk.}
    \scalebox{0.8}{
    \begin{tabular}{lccccccc}
      \hline
      Name & $M_{*}$ & $R_{*}$ & $L_{\nu,\mathrm{rad}}$ & $L_{X}$ & $L_{\mathrm{H}\alpha}$/$L_{\mathrm{bol}}$ &$T_{B,\textrm{plasma}}$ & Epoch \\
       & ($M_{\odot}$) & ($R_{\odot}$) & ($\times 10^{14}$\,erg\,s$^{-1}$\,Hz$^{-1}$) & ($\times 10^{28}$\,erg\,s$^{-1}$) & ($\times 10^{-4}$) & ($\times 10^{12}$\,K) & YYYY-MM-DD\\
       \hline
       \hline 
  DO\,Cep & 0.316 & 0.332 & 0.92$\pm$0.10 & 0.23 & 0.64 & 0.02 & 2019-02-07\\
  WX\,UMa & 0.095 & 0.121 & 0.45$\pm$0.09 & 0.36 & 1.21  & 0.04 & 2016-03-22\\
  AD\,Leo$^{*}$ & 0.420 & 0.431 & 0.96$\pm$0.10 & 3.20 & 1.72 & 0.15 & 2019-07-13\\
  GJ\,625$^{*}$ & 0.317 & 0.332 &0.80$\pm$0.09 & 0.04 & 0.06 & 0.004 & 2014-05-26\\
  GJ\,1151 & 0.167 & 0.190 & 0.63$\pm$0.15 & 0.02 &  0.07 & 0.005 & 2014-06-15 \\
  GJ\,450 & 0.460 & 0.474 & 0.54$\pm$0.20 & 0.66 & $<$0.09 & 0.04 & 2017-09-21 \\
  LP\,169-22 & 0.111 & 0.138 & 1.03$\pm$0.48 & $<$0.03 & & $<$0.05 &  2014-06-08\\
  CW\,UMa & 0.306 & 0.322 & 4.23$\pm$0.44 & 5.37 & 1.82 & 0.22 & 2019-09-23 \\
  HAT\,182-00605 & 0.442 & 0.422 & 2.94$\pm$0.57 & 3.40 &  & 0.13 & 2015-07-08\\
  LP\,212-62 & 0.161 & 0.183 & 28.34$\pm$1.55 & 0.38 & 0.84 & 0.03 & 2015-07-29\\
  DG\,CVn & 0.560 & 0.567 & 2.50$\pm$0.81& 10.72 & 3.03  & 0.41 & 2018-04-17 \\ 
  GJ\,3861 & 0.419 & 0.430 & 3.64$\pm$0.57 & 3.36 &  0.67 & 0.15 & 2017-04-19\\
  CR\,Dra & 0.823 & 0.827 & 43.38$\pm$2.46 & 36.65 & 1.38 & 0.91 & 2016-11-30 \\
  GJ\,3729 & 0.472 & 0.481 & 9.49$\pm$0.80 & 7.54 &  1.64 & 0.32 & 2015-01-21\\ 
  G\,240-45 & 0.125 & 0.162 & 12.30$\pm$1.57  & 0.02 &  & 0.01 & 2018-05-02\\
  2MASS\,J09481615+5114518 & 0.122 &0.161 & 28.71$\pm$2.27 & 0.28 &  & 0.03 & 2018-08-18\\
  LP\,259-39 & 0.173 & 0.202 & 10.11$\pm$2.75 & $<$18.70 & & 0.52 & 2018-06-11\\ 
  2MASS\,J10534129+5253040 & 0.408&0.379 & 26.39$\pm$2.55 &  28.01 & 3.74 & 0.30 & 2015-03-14\\
  2MASS\,J14333139+3417472 & 0.101&0.136 & 30.82$\pm$4.88 &  0.83 & 5.95 & 0.07 & 2015-10-10\\ 
    \hline
    \hline
    \end{tabular}}
\end{table*}

\subsection{Radio emission mechanisms}

There are several emission mechanisms that can produce the observed low-frequency radiation from M~dwarfs. The principal radiation characteristics that are used to differentiate between the plausible emission mechanisms are the observed brightness temperature, luminosity, degree of circular polarisation, duration, bandwidth, and coronal temperature of the star. These emission characteristics for our sample are provided in Table\,\ref{tab:radio_detections} and Supplementary Table\,\ref{tab:literature_info}.

In this section we only outline coherent emission mechanisms since all of our sources display a circularly polarised fraction $\gtrsim50\%$ and brightness temperature $\gtrsim$10$^{12}$\,K at 144\,MHz. Therefore, incoherent gyrosynchrotron radiation can be confidently excluded\cite{1985ARA&A..23..169D}. More detailed information about the emission physics can be found in Vedantham\cite{harish2020}.

\subsubsection{Plasma Emission}

Plasma emission results from the conversion of Langmuir waves to transverse electromagnetic waves via non-linear processes\cite{1985ARA&A..23..169D}. The commonly encountered mechanism that generates Langmuir waves is the bump-on-tail and loss-cone instabilities. In a stellar corona they can form when impulsively heated plasma ($T_{h} \sim 10^{8}\,K$) is injected into an ambient colder plasma ($T_{c} \sim 10^{6}\,K$). The impulsively heated plasma is a proxy for the characteristic velocity of the beam that emits Langmuir waves via the bump-in-tail instability. The resulting radio emission can occur at the fundamental plasma frequency and its second harmonic.

We use the expressions derived by Stepanov et al.\cite{Stepanov2001} to calculate the brightness temperature $T_{B,\textrm{plasma}}$ of the radio emission at the fundamental and the harmonic. The Langmuir wave spectrum was limited to wavenumbers ranging from those in resonance with the hot particles up to the Landau damping scale\cite{harish2019}. We also assume that the total energy density of the Langmuir waves is $10^{-5}$ of the kinetic energy of the ambient plasma. Such an assumption corresponds to the canonical level at which the turbulence alters the dispersion in the background plasma, leading to a runaway collapse of the Langmuir wave packets\cite{Benz1993,Reid2017}.
We estimate the ambient coronal temperatures for our detected stars via $T_{c} = 0.11F_{X}^{0.26}$, where $F_{X}$ is the X-ray surface flux calculated using a star's soft X-ray luminosity and the size of its photosphere\cite{Johnstone2015}. Such a relation has been shown to hold for both active and quiescent M~dwarfs. The coronal temperatures calculated for the X-ray detected stars range between $2 \times 10^{6}$\,K (GJ\,625) to $1 \times 10^{7}$\,K (LP\,259-39), with a mean of $5 \times 10^{6}$\,K. For LP\,169-22 and LP\,259-39, which are not detected in X-rays, we assumed an X-ray luminosity equal to the three-sigma upper limit derived from our \emph{XMM-Newton} observation and 2RXS, respectively. Finally, the hydrostatic density structure of the coronae of the stars was assumed to have a scale height\cite{harish2019} $L_{n} = 6 \times 10^{9} (T_{h}/10^{6}\,\mathrm{K})(R_{*}/R_{\odot})^{2}(M_{*}/M_{\odot})^{-1}$. 

To calculate the brightness temperature of any potential plasma emission from our detected stars, we used Equations 15 to 22 of Stepanov et al.\cite{Stepanov2001} and varied the temperature of the injected hot plasma to be up to 30 times the coronal temperature. Such temperature ranges are consistent with the observed temperatures of coronal loops\cite{Giampapa1996} and flares\cite{Robrade2010} on M~dwarfs. In general, the plasma temperatures of M~dwarf flares are less than 30 MK\cite{Robrade2010,Stepanov2001,Osten2006}.

To make the comparison fair to the observed brightness temperatures reported in Table\,\ref{tab:radio_detections}, we assume the size of the emitter is the stellar disk. Such an assumption is likely accurate to at least within a factor of two, as plasma source regions observed on the Sun generally do not exceed the size of the Solar photometric disk\cite{Saint-Hilaire2013}. The estimated brightness temperature of fundamental plasma emission $T_{B,\mathrm{plasma}}$, when the injected heated plasma is equal to 30 times the coronal temperature, is provided in Supplementary Table\,\ref{tab:literature_info}.

We note that the assumptions required in this calculation imply that the estimated brightness temperature is only accurate to an order of magnitude for most of the sources. This is largely because it is unclear what is the total energy density of the Langmuir waves for each star. If we allow extreme values for the energy density, such as $\approx 10^{-4}$, fundamental plasma emission can produce the observed brightness temperature for about half of the sample. However, even in this extreme case, the plasma emission from the fundamental cannot achieve the measured brightness temperatures for our best quiescent  candidates (GJ\,625, GJ\,450, LP\,169-22, G\,240-45, and GJ\,1151) and more distant systems (LP\,212-62, 2MASS\,J10534129+5253040, and 2MASS\,J14333139+3417472).

Plasma emission from the second harmonic can reach the observed brightness temperatures for all our detections. However, harmonic emission can not generate the high circularly polarised fraction detected for our sample. For example, Solar plasma harmonic emission has only been observed to have a circular polarised fraction $<20\%$\cite{Benz1993}. For the most extreme configurations, where coronal loops occupy the entire stellar disk, the circularly polarised fraction of harmonic plasma emission is limited to $\approx$60\%\cite{Melrose1978,Melrose1980,Ledenev1994}. Only our detections of AD\,Leo and DO\,Cep have circularly polarised fractions $<60\%$. Furthermore, even if the entire stellar disk was occupied by coronal loops, it is likely that the emission from regions with opposite magnetic field orientation would result in substantially lower circularly polarised fraction.

Therefore, we argue that plasma emission from the fundamental can plausibly generate the high observed brightness temperature and circularly polarised fraction for some, but not all, of our detections. Only the observed emission from AD\,Leo and DO\,Cep are potentially consistent with harmonic plasma emission due to their low circular polarised fractions. 

We caution that the theoretical plausibility of plasma emission does not mean that it is the true mechanism. Plasma emission from flare stars is usually attributed to flaring coronal loops close to the surface\cite{1983SoPh...88..297Z} whereas, in our brightness temperature calculation, we have implicitly assumed that the whole surface of the star is filled with loops that are simultaneously flaring. There is no precedent for such a scenario. In addition, induced emission at the fundamental is significantly more efficient at higher frequencies ($\nu\sim 1\,{\rm GHz}$). Therefore, if disc-averaged $T_b\sim 10^{12}-10^{13}\,{\rm K}$ from our sample is attributed to fundamental plasma emission, then we would expect to long-lived emission with $T_b\gg 10^{13}\,{\rm K}$ at $\nu\gtrsim 1\,{\rm GHz}$. Such emission has not been observed even on the most active stars\cite{2019ApJ...871..214V}. Finally, if we are indeed observing plasma emission from short flaring loops, then we expect the polarity of the emission to change over time as loops with different orientation with respect to the line of sight come in and out of view. We do not observe this. For example, all 20 of CR\,Dra's low-frequency radio detections have a constant circularly-polarised handedness. The constant positive handedness of the circularly polarised light over 6.5 days of monitoring, taken in two observing blocks separated by a year\cite{callinghamCRDra}. For all our stars in which we have multiple detections, no star has ever flipped polarity (with follow-up observations occurring weeks and up to three years apart).

The formalism presented above also allows us to place some key coherent and low-frequency detections in the context of our detections. For example, Slee et al.\cite{Slee2003} detected coherent, narrowband emission from Proxima Centauri that persisted for over two-days at $\sim$\,1.4\,GHz. While the emission was 100\% circularly polarised, the detected brightness temperature of $3\times 10^{9}$\,K is over three orders of magnitude lower than the brightness temperatures we measure for the majority of our sources. An application of the equations in Stepanov et al.\cite{Stepanov2001} shows that the observed brightness temperature could readily be supplied by plasma emission. Additionally, Lynch et al.\cite{2017ApJ...836L..30L} recently detected coherent emission from the flare star UV\,Ceti at 154\,MHz with the Murchison Widefield Array (MWA). While the emission achieved brightness temperatures of $10^{13}- 10^{14}$\,K, the non-detection in total intensity only provided a lower limit to the circularly polarised fraction of $\gtrsim 27\%$. Such emission is therefore also consistent with harmonic plasma emission, similar to our detections of AD\,Leo and DO\,Cep. However, we note that the MWA detections of UV\,Ceti appear consistent with flares lasting $\sim$30\,min, while AD\,Leo and DO\,Cep have persistent emission lasting $\gtrsim$\,4\,hrs.

\subsubsection{Electron-cyclotron maser emission}

For low density, highly magnetised plasmas, coherent radio emission can be produced via the electron cyclotron maser instability (ECMI). There are several possible astrophysical configurations that can produce the population inversion necessary for the classical maser to operate. One common configuration is the unstable loss-cone distribution of electron energies set up in a stellar coronal loops\cite{1985ARA&A..23..169D}. In this case, the brightness temperature of a continuously operating maser is\cite{melrose1982,harish2019}

\begin{equation}
T_{B,\mathrm{flare}} \approx 2 \times 10^{14}(\lambda/200\,\mathrm{cm})^{2}(L/R_{*})^{-1}(\beta/0.2)^{2}\,\mathrm{K},
\label{eqn:coronal_loop}
\end{equation}

\noindent where $\lambda$ is the wavelength of radiation, $R_{*}$ is the stellar radius, $L$ is the length-scale of the magnetic trap, and $\beta = v/c$, corresponding to the velocity of the electrons $v$ divided by the speed of light $c$. ECMI emission will occur at a frequency dependent on the opening angle of the loss-cone and energy of the electrons. While the bandwidth of a single pulse of the maser will be small, the radiation will occur at a range of heights in the flux tube, generating broadband emission as high as 4:1\cite{1985ARA&A..23..169D}.

To check if the broadband emission of the observed brightness temperature can be generated by a compact coronal loop, we consider a heuristic model of a coronal loop of length $L$ and a characteristic width of $W=0.1L$ \cite{harish2019}. The instantaneous bandwidth of the loss cone maser is $\Delta\nu/\nu\approx \beta^2\alpha^2$ where $\alpha$ is the opening angle of the loss cone\cite{melrose1982}. Hence, to produce the observed broadband emission, the loop can be conceptually thought of as $\nu/\Delta\nu$ maser sites each radiating with a brightness temperature given by Equation\,\ref{eqn:coronal_loop}. For an observed flux density of $1\,{\rm mJy}$ at the distance of $20\,{\rm pc}$ (the median distance to our population), we arrive at a rough constraint on the loop length of $L\sim 25R_\ast$ for characteristics values of $\beta=\alpha=0.5$. This shows that a loss cone maser operating in a compact coronal loop ($L\lesssim R_\ast$) cannot account for the observed emission. Even in cases of the most active M~dwarfs, coronal loops have only been observed to extend to $\sim 4$ stellar radii\cite{Benz1998}. We note that we could also have modelled this situation as many small-scale loops participating in the maser emission. However, that would likely result in a low circular polarisation fraction, or changing handedness of the polarisation. Additionally, if the observed emission was generated from coronal loops, we would have expected correlations with other coronal activity indicators, such as X-rays. Therefore, the observed emission is likely generated within a larger global magnetospheric trap. 

As elaborated in the main text, an alternative way to drive ECMI radiation is through the breakdown of co-rotation. For example, the volcanic activity of Io in the Jovian system produces an equatorial plasma sheet that rotates with Jupiter. The plasma begins to lag behind the magnetic field of Jupiter at a radius of $\approx$6 Jovian radii, generating a magnetosphere–ionosphere coupling current that accelerates electrons into the neutral atmosphere of Jupiter\cite{Hill1976}. The deposition of high-energy electrons at the poles of Jupiter set-up the population inversion necessary for ECMI to occur\cite{Cowley2001}. A similar particle system can also be established for M~dwarfs, where continual flaring behaviour could deposit a torus of plasma around the star or, potentially, a volcanically active satellite.

Following the formulation of Nichols et al.\cite{2012ApJ...760...59N}, ECMI radiation generated by the breakdown of co-rotation can only achieve the observed radio spectral luminosities (Supplementary Table\,\ref{tab:literature_info}) for stellar rotation periods $\lesssim$2\,days. Excluding the systems in which we think plasma emission is a plausible origin of the low-frequency radiation, and stars without known rotation periods, the breakdown of co-rotation model can generate the observed radio luminosities for WX\,UMa, DG\,CVn, CR\,Dra, and (potentially) HAT\,182-00605. 

The physics of such systems share a similarity to the detection of ECMI radiation from rapidly rotating ultracool dwarfs\cite{2008ApJ...684..644H} and V374 Peg\cite{2018ApJ...854....7L}. Additionally, the linearly-polarised $\approx$1\,GHz bursts identified by Zic et al.\cite{Zic2019} from UV\,Ceti are likely produced by ECMI from the breakdown of co-rotation. However, the bursts that can be confidently associated with ECMI by Zic et al.\cite{Zic2019} appear to last $\lesssim$\,20\,min and spaced at least several hours apart. In comparison, the majority of our sample are observed to emit highly-circularly polarised light for at least eight hours. This may imply that we have a preferential line of sight to the emission site(s).

Finally, if breakdown of co-rotation was the sole mechanism driving the coherent radiation emission in our sample, we would expect rapidly-rotating systems to be preferentially detected. We show in Supplementary Information Section ``Rotation period and detection rate'' that that is not the case.

An alternative mechanism that can drive ECMI radiation is a Sub-Alfv\'{e}nic star--planet interaction. Expressions for the radio power generated by a sub-Alfv\'{e}nic interaction of a star and its exoplanet have been originally provided by Zarka et al.\cite{2007P&SS...55..598Z}. Here we use the slightly modified expressions of Saur et al.\cite{2013A&A...552A.119S} and Turnpenney et al.\cite{2018ApJ...854...72T} to compute the starward energy flux from a sub-Alfv\'{e}nic interaction with an exoplanet:
\begin{equation}
    S^{\rm th} = \frac{\alpha^2}{2}\sin^2\theta v_{\rm rel}^2B R_{\rm eff}^2\sqrt{4\pi \rho},
    \label{eqn:star_eng_flux}
\end{equation}
in Gaussian units, where $\alpha$ is the interaction strength\cite{2013A&A...552A.119S}, $v_{\rm rel}$ is the velocity of the stellar wind flow in the frame of the planet, $B$ and $\rho$ are the stellar magnetic field strength and wind density at the location of the planet, and $R_{\rm eff}$ is the effective radius of the planetary obstacle to the stellar wind flow. The starward Poynting flux is converted to ECMI radiation with an efficiency $\epsilon_{rad}$. The flux density of the observed emission is then 
\begin{equation}
F_{\rm rad} = S\epsilon_{\rm rad}/(\Omega d^2 \Delta\nu),
\label{eqn:flux}
\end{equation}
where $\Omega$ is the beam solid angle of the emitter, $d$ is the distance to the stellar system, and $\Delta\nu$ is the total bandwidth of emission.  Because the emission has a cut-off at a frequency corresponding to the cyclotron frequency at the stellar surface, the bandwidth of emission can be approximated as $\Delta\nu=eB_\ast/(2\pi m_e) $ where $B_\ast$ is the surface field strength of the star, $e$ and $m_e$ are the electron charge and mass respectively. Substituting for the starward energy flux, we get 
\begin{equation}
    F_{\rm rad} = \frac{\pi m_e}{e}\frac{\alpha^2\epsilon_{\rm rad}}{d^2\Omega}R_{\rm eff}^2\sin^2\theta v_{\rm rel}^2 \sqrt{4\pi \rho} \frac{B}{B_\ast}.
    \label{eqn:flux_rad}
\end{equation}
We note that because $B$ is proportional to $B_\ast$, to first order, the radio flux density is largely insensitive to the stellar magnetic field, although the frequency band over which a particular star radiates depends on the stellar magnetic field strength. This is consistent with the lack of dependence of the radio luminosity seen in our sample with proxies for the stellar magnetic field strength such as rotation rate and/or Rossby number.

We examined the feasibility of the exoplanet-induced model in terms of the energy requirements as follows. We adopt a radiation efficiency of $\epsilon_{\rm rad}=0.01$, a conservative estimate\cite{2012ApJ...760...59N} as numerical simulations allow for $\epsilon_{\rm rad}\approx 0.1$. We assume a beam-solid angle for the emission of $\Omega = 0.15$ as expected for an electron speed of $0.2c$. The starward Poynting flux inferred from observations can then be computed using equation\,\ref{eqn:flux} as 
\begin{equation}
    S^{\rm obs} \approx 4\times 10^{22}\,\,\left(\frac{F_{\rm rad}}{\rm mJy}\right)\left(\frac{d}{10\,{\rm pc}}\right)^2 \left(\frac{\Omega}{0.15\,{\rm sr}}\right)\left(\frac{B_\ast}{100\,{\rm G}}\right)\,\,\,{\rm erg/s}.
\end{equation}

The theoretically expected Poynting flux can be computed from Equation\,\ref{eqn:star_eng_flux}. We adopt $v_{\rm rel} = 500\,{\rm km/s}$ which is typical of winds driven by a thermal expansion of $\sim 10^6\,{\rm K}$ coronae. We computed the density $\rho$ by conservatively adopting a base coronal density of $n_{\rm b} = 10^7\,{\rm cm}^{-3}$, a value that is an order of magnitude lower than the solar value, and assuming that radial expansion leads to a $n_{\rm b}r^{-2}$ fall-off in density with distance $r$ from the star in units of the stellar radius. We assume a Parker spiral configuration for the magnetic field with a surface field strength of $B_\ast = 100\,{\rm G}$. We emphasise that the Alfv\'{e}nic point only depends on the radial structure of the field strength and plasma density, rather than the azimuthal structure of the field. The radial profile of the angle $\theta$ in Equation\,\ref{eqn:star_eng_flux} depends on the rotation rate of the star, the orbital period of the planet and the wind speed all of which will change substantially from system to system. As a benchmark value, we adopt $\theta = 0.2$ which corresponds to the case of a planet orbiting at a distance of $r=10$  (units of stellar radii) around a star with a 500\,km/s wind and rotation period of 10\,days. We assume a terrestrial Earth-like planet with $R_{\rm eff} = 6500$ km. Finally, we assume $\alpha=1$, which corresponds to the case of a planet with a highly conductive atmosphere (see Equation 17 of Saur et al.\cite{2013A&A...552A.119S}). The starward Poynting flux for our benchmark case is then
\begin{equation}
    S^{\rm th} \approx  6\times 10^{22}\,\left(\frac{\sin\theta}{0.2}\right)^2 \left(\frac{v_{\rm rel}}{500\,{\rm km/s}}\right)^2 \left(\frac{B_\ast}{100\,{\rm G}}\right) \left( \frac{R_{\rm eff}}{6500\,{\rm km}}\right)^2\left(\frac{n_{\rm b}}{10^7\,{\rm cm}^{-3}}\right)^{1/2} \left(\frac{r}{10}\right)^{-3}\,\,\,{\rm erg/s}.
\end{equation}

Applying the aforementioned values to this equation allows us to recover the radio flux densities and luminosities listed in Table\,\ref{tab:radio_detections} and Supplementary Table\,\ref{tab:literature_info}. Therefore, ECMI radiation driven by a Sub-Alfv\'{e}nic flux from star--planet interaction can produce the appropriate luminosity, duration, bandwidth, and brightness temperature for our sample, especially since we observe no correlation with other stellar activity markers. The model is particularly well suited to explain the origin of bright radio emission from sources that before to this study, and that of Vedantham et al.\cite{harish2019}, would not have been considered potential radio-emitters, such as GJ\,625, GJ\,450, GJ\,1151, LP\,169-22, G\,240-45, and LP\,212-62.

Finally, since we argue it is plausible that breakdown of co-rotation and star-planet interactions are driving the radio emission for some of our detections, beaming effects may substantially vary the flux density of our detections epoch to epoch. However, such intrinsic variations does not necessarily preclude observable correlations because stars with all activity levels are equally affected by beaming. In other words, the intrinsic variations affect quiescent and active stars in the same way so underlying correlations (or lack thereof) are not destroyed.

\subsection{G{\"u}del-Benz relationship}

\begin{figure*}[h]
\renewcommand\figurename{Supplementary Information Figure}
   \centerline{\resizebox{0.7\textwidth}{!}{\includegraphics{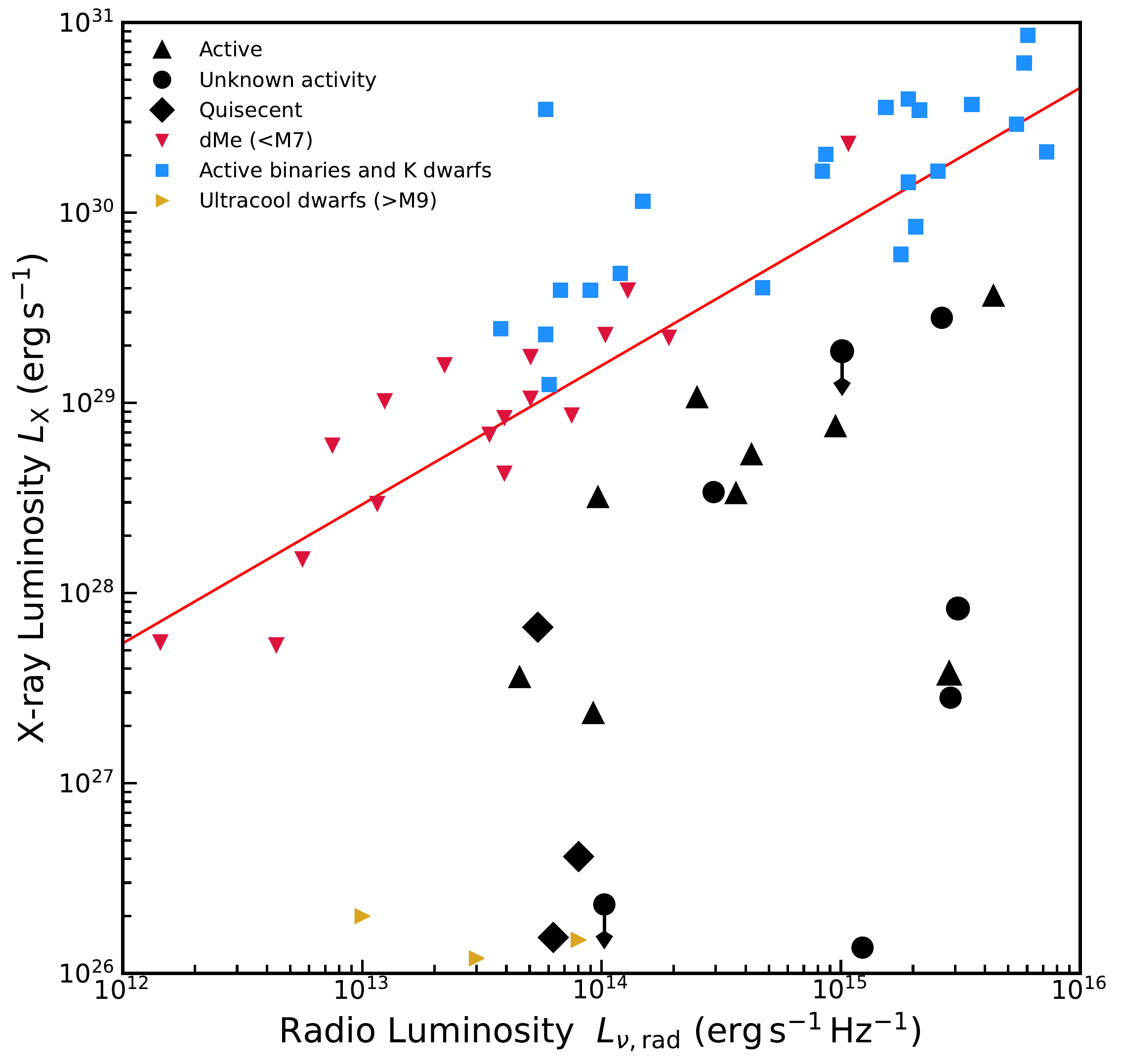}}}
 \caption{\label{fig:gb} 
 {\bf Soft (0.2-2.0\,keV) X-ray luminosity $L_{\mathrm{X}}$ against radio luminosity $L_{\nu,\mathrm{rad}}$.} 
 The literature data for the chromospherically-active stars used to derive the original G{\"u}del-Benz relation are plotted as coloured triangles or squares\cite{1993ApJ...405L..63G,1994A&A...285..621B}, with the best fit to the literature data indicated by the red line ($L_{\mathrm{X}} \propto L_{\nu,\mathrm{rad}}^{0.73}$). Some ultracool dwarfs with X-ray and radio detections are shown in yellow\cite{2014ApJ...785....9W}. There are other ultracool dwarf systems with lower X-ray luminosities that are off the scale of this plot. Our 144\,MHz radio sample (all with spectral types earlier than M7) is represented by black symbols, with triangles, diamonds, and circles indicating chromospherically active, quiescent, and unknown chromospheric activity, respectively. The chromospheric activity classifications are derived from H$\alpha$ measurements (see Supplementary Information Figure\,\ref{fig:halpha}). Non-detected X-ray sources are shown as 3-$\sigma$ upper limits. While the radio luminosities plotted for the literature population are derived from 5 to 8\,GHz observations, all of our detections have flat spectra.}

\end{figure*}

One test of whether our sample belongs to the class of canonical radio stars characterised by gigahertz-frequency observations\cite{Gudel2002} is to compare 144\,MHz radio $L_{\nu,\mathrm{rad}}$ and 0.2-2.0\,keV X-ray $L_{\mathrm{X}}$ luminosity, as shown in Supplementary Information Figure \ref{fig:gb}. The correlation between quiescent gyrosynchrotron radio and soft X-ray luminosity, referred to as the G{\"u}del-Benz relation where $L_{\mathrm{X}} \propto L_{\nu,\mathrm{rad}}^{0.73}$, holds over 10 orders of magnitude in radio luminosity despite the fundamentally different emission mechanisms for the radio and soft X-ray radiation\cite{1993ApJ...405L..63G,1994A&A...285..621B}. The standard model used to explain the G{\"u}del-Benz relation posits that magnetic reconnection events in the stellar corona produce a population of non-thermal electrons that emit in the radio. This non-thermal population deposits some of its energy in the chromosphere of the star, producing an enhancement in soft X-rays as some of the newly ablated material concentrates in coronal loops\cite{1994A&A...285..621B,2014ApJ...785....9W}. 

Since we know that a coherent emission mechanism is operating in our detected stars, we expect the sample should deviate from the established G{\"u}del-Benz relation. As observed in Supplementary Information Figure\,\ref{fig:gb}, our sample significantly deviates from the established G{\"u}del-Benz relation, with some stellar systems (e.g. GJ\,1151, GJ\,625, LP\,212-62) departing from the relation by two to three orders of magnitudes. While it is not surprising that coherent auroral radio emission does not follow this canonical relationship\cite{2008ApJ...684..644H}, it is noteworthy that the observed low-frequency radio luminosity does not show a clear dependence on the coronal temperature and density traced by the X-ray luminosity. For example, the seven detected stars at similar radio luminosities around $\sim 10^{14}\,{\rm erg\,s}^{-1}\,{\rm Hz}^{-1}$ display a spread of over two orders of magnitude in their X-ray luminosity. Additionally, the systems do not establish a clear new relationship, suggesting that chromospheric activity has limited influence on the detected radio luminosity.

Such an inference is also supported by the constant low-frequency detection rate from spectral types M1.5 to M6.0, representing substantially different coronal temperatures and dynamo mechanisms. The detection of spectral types from M1.5 to M6.0 also implies that the reduction of coronal plasma heating efficiency can not resolve the violation of the canonical G{\"u}del-Benz relation in our low-frequency detected population, as sometimes used to explain why ultra-cool dwarfs deviate from the G{\"u}del-Benz relation\cite{2002ApJ...572..503B,2010ApJ...709..332B}.

Finally, we note that there are underlying uncertainties when comparing our detections to the literature G{\"u}del-Benz relation. The relation was established by observations between 5 and 8 GHz, while our observing frequency is significantly lower. Therefore, any spectral structure will impact the comparison.

\newpage

\subsection{Variability}

\begin{figure*}
\renewcommand\figurename{Supplementary Information Figure}
   \centerline{\resizebox{\textwidth}{!}{\includegraphics{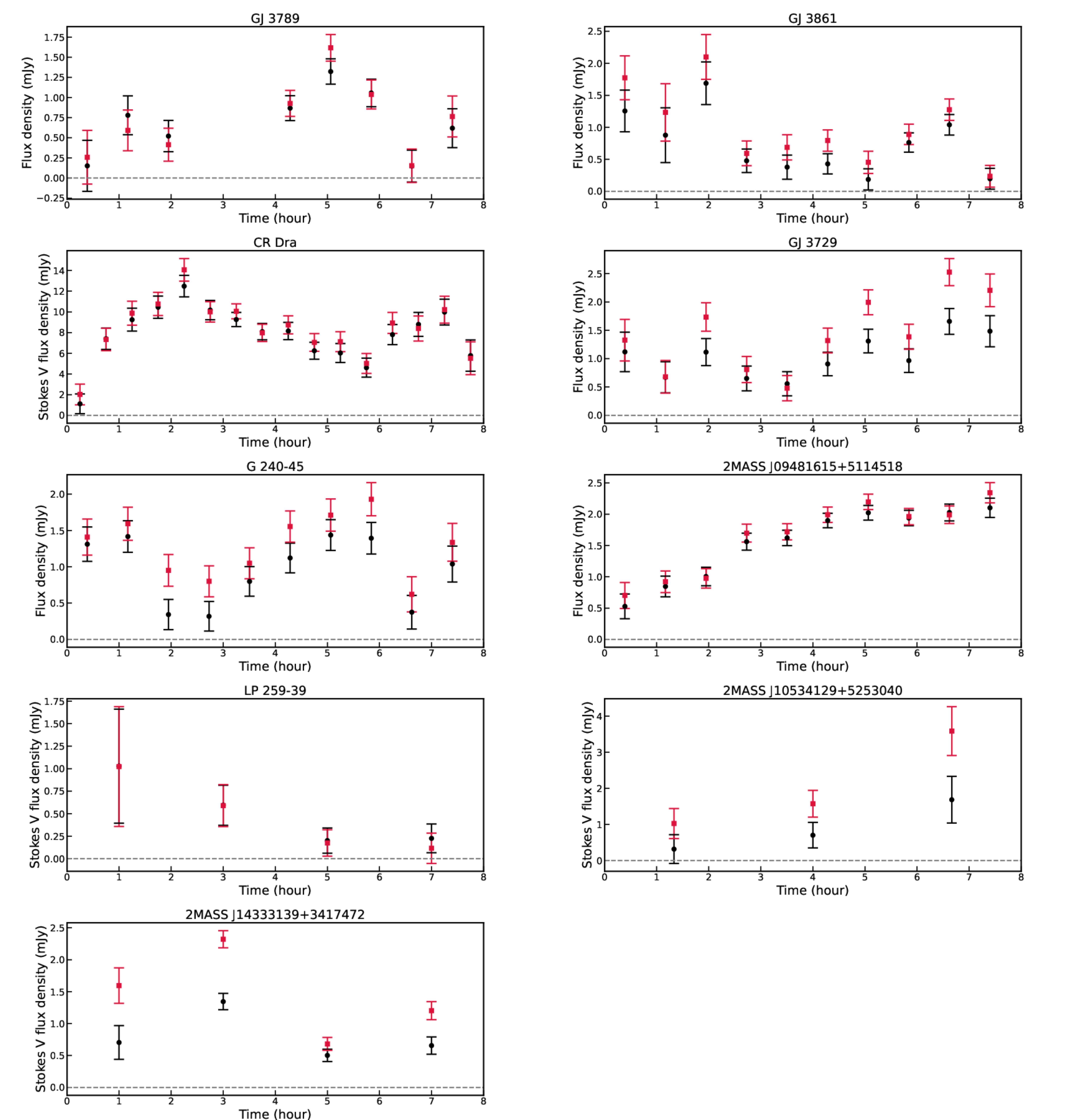}}}
 \caption{\label{fig:lightcurves2} 
 {\bf 144\,MHz lightcurves for the detections listed in Table 1 (continued).} Total intensity and absolute circular polarisation measurements are represented by red squares and black circles, respectively. To infer the sign of the circular polarisation, refer to Table\,\ref{tab:radio_detections}. The title of each panel corresponds to the name of the stellar system. The dashed grey line represents zero level. Any measurement consistent with zero should be considered a non-detection. The error bars represent 1-$\sigma$ uncertainty.}

\end{figure*}

The detection strategy of our survey involves searching LOFAR fields that have been synthesised over 8\,hours of observation. Therefore, our survey can be biased towards detecting sources that show large, short duration bursts, or relatively constant emission that continues for multiple hours. We are confident that the emission we detect from our sources are of the latter type for two reasons. Firstly, if the sources were emitting bright ($>10$\,mJy) short ($<1$\,hr) bursts, they would have been previously detected by similar lower-sensitivity all-sky low-frequency Stokes~V searches\cite{2018MNRAS.478.2835L}. Secondly, we can form 144\,MHz lightcurves for our sources, provided we integrate over the available bandwidth. The lightcurves for all our detected sources are shown in Supplementary Information Figures\,\ref{fig:lightcurves} and \ref{fig:lightcurves2}, where we have presented the absolute value of Stokes\,V emission. The sign of the circularly polarised emission can be inferred from Table\,\ref{tab:radio_detections}.

\begin{figure*}
\renewcommand\figurename{Supplementary Information Figure}
   \centerline{\resizebox{\textwidth}{!}{\includegraphics{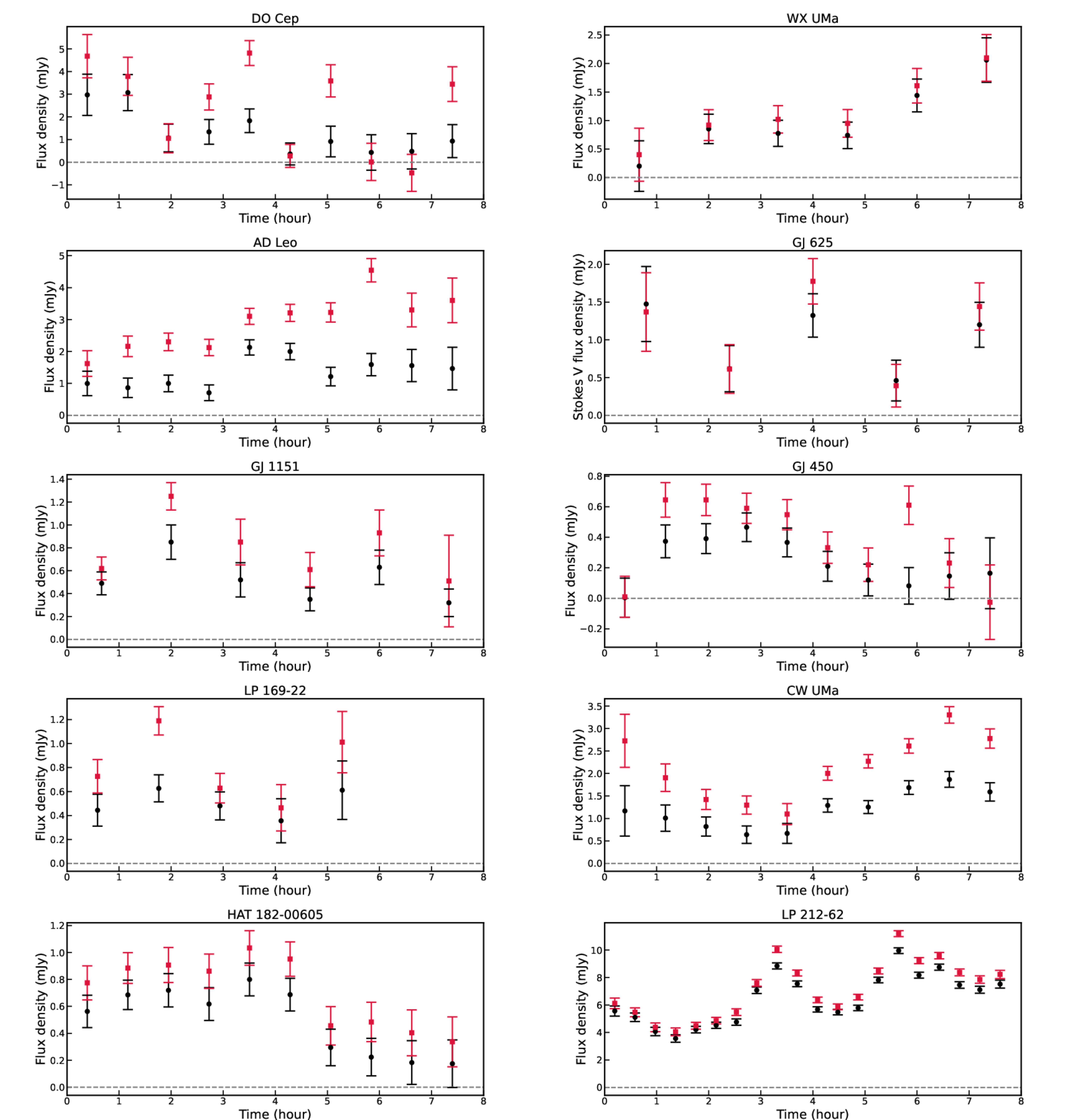}}}
 \caption{\label{fig:lightcurves} 
 {\bf 144\,MHz lightcurves for the detections listed in Table 1.} Total intensity and absolute circular polarisation measurements are represented by red squares and black circles, respectively. To infer the sign of the circular polarisation, refer to Table\,\ref{tab:radio_detections}. The title of each panel corresponds to the name of the stellar system. The dashed grey line represents zero level. Any measurement consistent with zero should be considered a non-detection. The error bars represent 1-$\sigma$ uncertainty.}

\end{figure*}

For all of our detected sources we can sample the lightcurves in three or more time bins. The flux density for each time sample was measured from images formed from integrating over 48\,MHz bandwidth, and using the position of the star as a prior in forced-photometry fitting. The uncertainties represent 1-$\sigma$, and any uncertainty consistent with zero should be considered a non-detection. Different signal-to-noise properties influenced the variable sampling of the lightcurves, particularly the position of the source relative to the observing pointing centre or any non-Gaussian noise properties. Note that the data for the last $\approx20\%$ of LP\,169-22's and middle of GJ\,3729's observation are not included as the data was heavily impacted by radio frequency interference.

All of the lightcurves show highly circularly polarised emission that lasts at least 4 hours, and in the majority of cases, the entirety of the 8 hours of observation. Furthermore, no lightcurve shows characteristics of flare-like behaviour or a statistically significant flip in the handedness of the circular polarisation. We also made lightcurves with time spans of minutes to tens of minutes, but no stellar system showed temporal structure expected from Type\,II or III bursts. The lightcurves and the spectra of the detected sources will be studied in detail in a follow-up manuscript.

Finally, the variability of our sources between observing epochs, as listed in Table\,\ref{tab:radio_detections}, could be used to test the favoured emission mechanism. For example, ECMI radiation will be beamed along the surface of a cone that will be modulated by the rotation of the star or the orbit of a putative satellite. However, the sampling currently available for most our stellar systems is not adequate to provide meaningful constraints considering the number of degrees of freedom in the radiation model, such as magnetic field and orbital orientation. Additionally, we note that our observing strategy is naturally biased to systems in which the geometry of the emission is preferential to our line of sight.

\subsection{Rotation period and detection rate}

Previous gigahertz-frequency studies of M and ultracool dwarfs have shown a clear increase in the fraction of radio detected stars with rotation period\cite{2012ApJ...746...23M} -- a star with less than $\sim$2 days rotation period is at least five times more likely to be detected at gigahertz-frequencies than a star rotating with period greater than $\sim$2 days. The relationship between rotation rate and radio detection is used as evidence that stellar rotation regulates the magnetic field strength and topology in the corona via the stellar dynamo\cite{Gudel2002,2012ApJ...746...23M}. 

If the 144\,MHz stellar system emission that we detect is driven by an amalgamation of breakdown of corotation and star-planet interactions, we would not expect a strong dependence of detection fraction with stellar rotation. While stars with rotation $\lesssim2$\,day will be able to generate coherent low-frequency radio emission from the breakdown of corotation, stars with longer rotational periods far outnumber such rapid rotators, and these more plausibly generate their emission from a star-planet interaction. In Supplementary Information Figure\,\ref{fig:det_rot} we show our 144\,MHz detection rate with rotation period for all the M~dwarfs reported by Newton et al.\cite{2017ApJ...834...85N} that were in our survey footprint and at a distance less than 33\,pc, the limiting distance of the study by Newton et al.\cite{2017ApJ...834...85N}. We do not observe a trend with rotation period. While this figure is impacted by small number statistics, the number of detections implies we are sensitive to a change in detection rate of a factor of five or more. 

\begin{figure*}
\renewcommand\figurename{Supplementary Information Figure}
\centerline{\resizebox{0.6\textwidth}{!}{\includegraphics{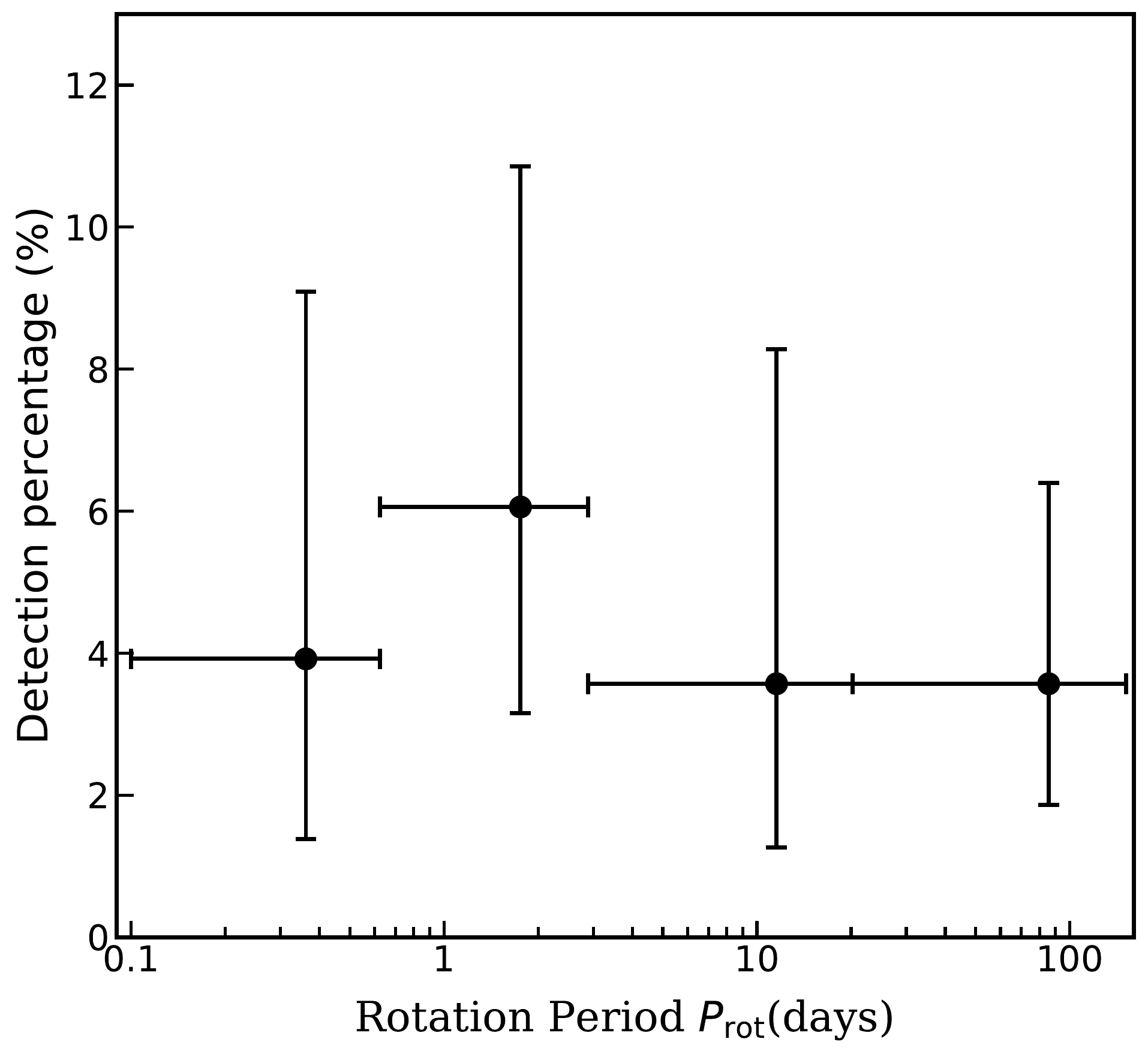}}}
 \caption{{\bf 144\,MHz detection rate as a function of rotation period for the M~dwarfs studied by Newton et al.\cite{2017ApJ...834...85N}.} While the sample produced by Newton et al.\cite{2017ApJ...834...85N} is inhomogeneous and incomplete at periods $\gtrsim80$\,days, the lack of a trend in the detection rate with rotation is in contrast to what is observed by gigahertz-frequency studies\cite{2012ApJ...746...23M}. The lack of an observed trend supports the conclusion that the 144\,MHz emission being driven by a combination of breakdown of co-rotation and star-planet interactions. Due to incompleteness in the Newton et al.\cite{2017ApJ...834...85N} sample, the absolute detection percentages should not be extrapolated to other surveys. The error bars represent 1-$\sigma$ uncertainty.}
\label{fig:det_rot}
\end{figure*}

We note that there are several important caveats to Supplementary Information Figure\,\ref{fig:det_rot}. In particular, while the work of Newton et al.\cite{2017ApJ...834...85N} is the most comprehensive study of rotation in M~dwarfs to date, it is inhomogeneously formed and almost certainly incomplete for stars with rotation periods $\gtrsim80$\,days. However, the lack of an observed trend between 0.1 and 80 days is robust to incompleteness. This incompleteness and inhomogeneity also implies that the absolute value of the detection rate should not be extrapolated to other surveys.

\subsection{H$\alpha$ characteristics of the population}

Chromospheric H$\alpha$ emission is a tracer of magnetic activity, and is known to have a have a strong relationship with rotation in M~dwarfs\cite{2017ApJ...834...85N}. If the radio emission we have detected is related to the underlying dynamo, we could expect a relationship with H$\alpha$ emission. As shown in Supplementary Information Figure\,\ref{fig:halpha}, we see no correlation with H$\alpha$ normalised by bolometric luminosity and 144\,MHz radio luminosity $-$ sources with equivalent H$\alpha$ emission can vary by nearly 2 orders of magnitude in radio luminosity. Also, our sample spans the whole chromospheric activity range from non-detections in H$\alpha$ to stars with H$\alpha$ equivalent widths of $\sim$10\,\AA.

\begin{figure*}
\renewcommand\figurename{Supplementary Information Figure}
\centerline{\resizebox{\textwidth}{!}{\includegraphics{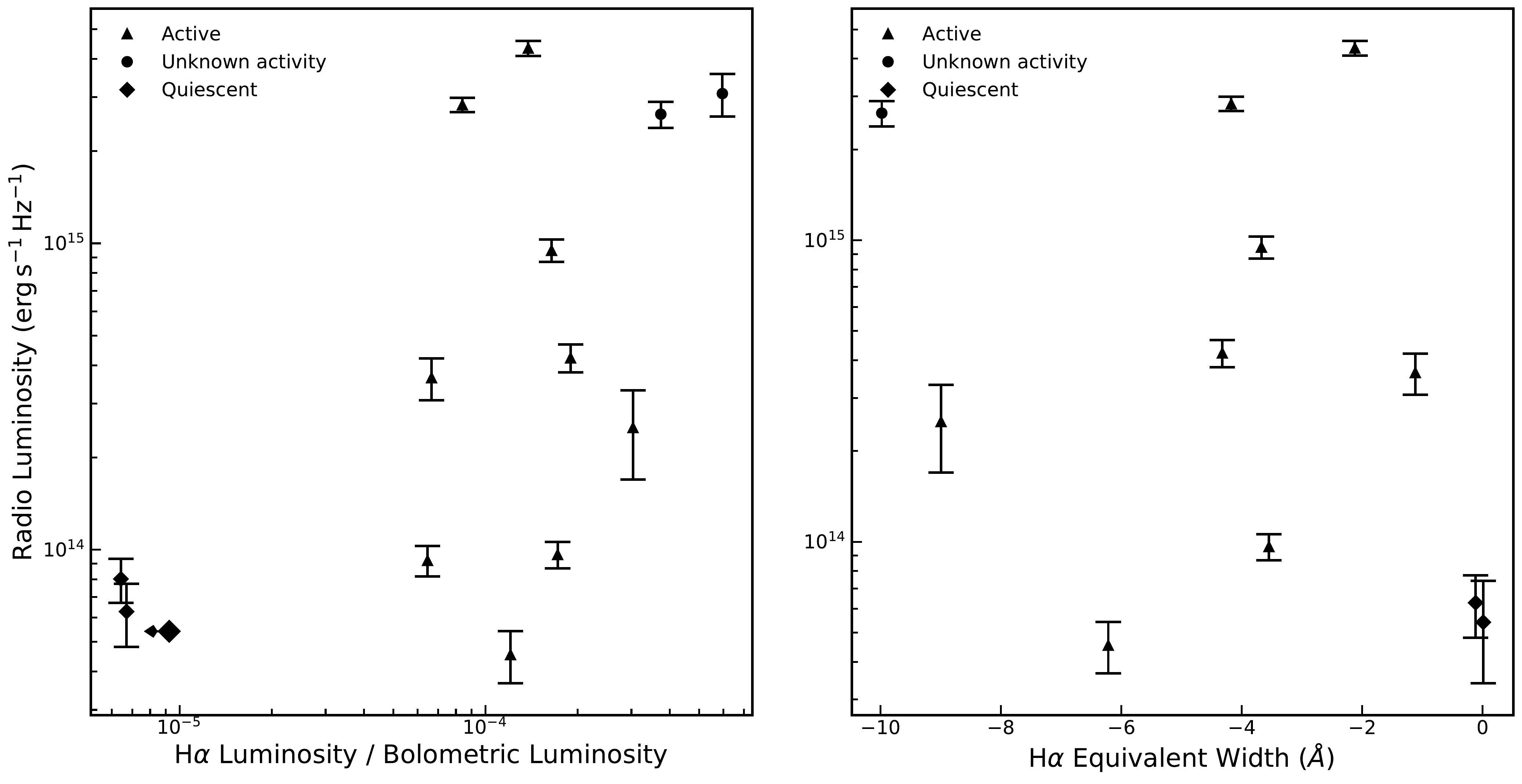}}}
 \caption{{\bf 144\,MHz radio luminosity against H$\alpha$ luminosity by bolometric luminosity (left panel) and H$\alpha$ equivalent width (right panel).} The ratio of H$\alpha$ luminosity to bolometric is relative to ``inactive'' stars defined by Newton et al.\cite{2017ApJ...834...85N}. Upper-limits and error bars represent 1-$\sigma$ uncertainty.}
\label{fig:halpha}
\end{figure*}

\subsection{144\,MHz source counts of M~dwarf stellar systems}

Radio source counts $N$ of a flux density-limited $S$ sample can provide information about survey completeness and an estimate of the number of detections that a more sensitive survey could achieve\cite{1987IAUS..124..545K}. The Stokes~V and Stokes~I differential source counts for our 19 M\,dwarf stellar systems detected at 144\,MHz in LoTSS fields that were observed for 8\,hours are presented in Supplementary Information Figure\,\ref{fig:source_counts}. Despite the sources being of Galactic origin, we perform a Euclidean Universe normalisation by multiplying the differential counts by $S^{2.5}$ to be consistent with previously presented radio source counts\cite{1987IAUS..124..545K}.

\begin{figure*}
\renewcommand\figurename{Supplementary Information Figure}
\centerline{\resizebox{\textwidth}{!}{\includegraphics{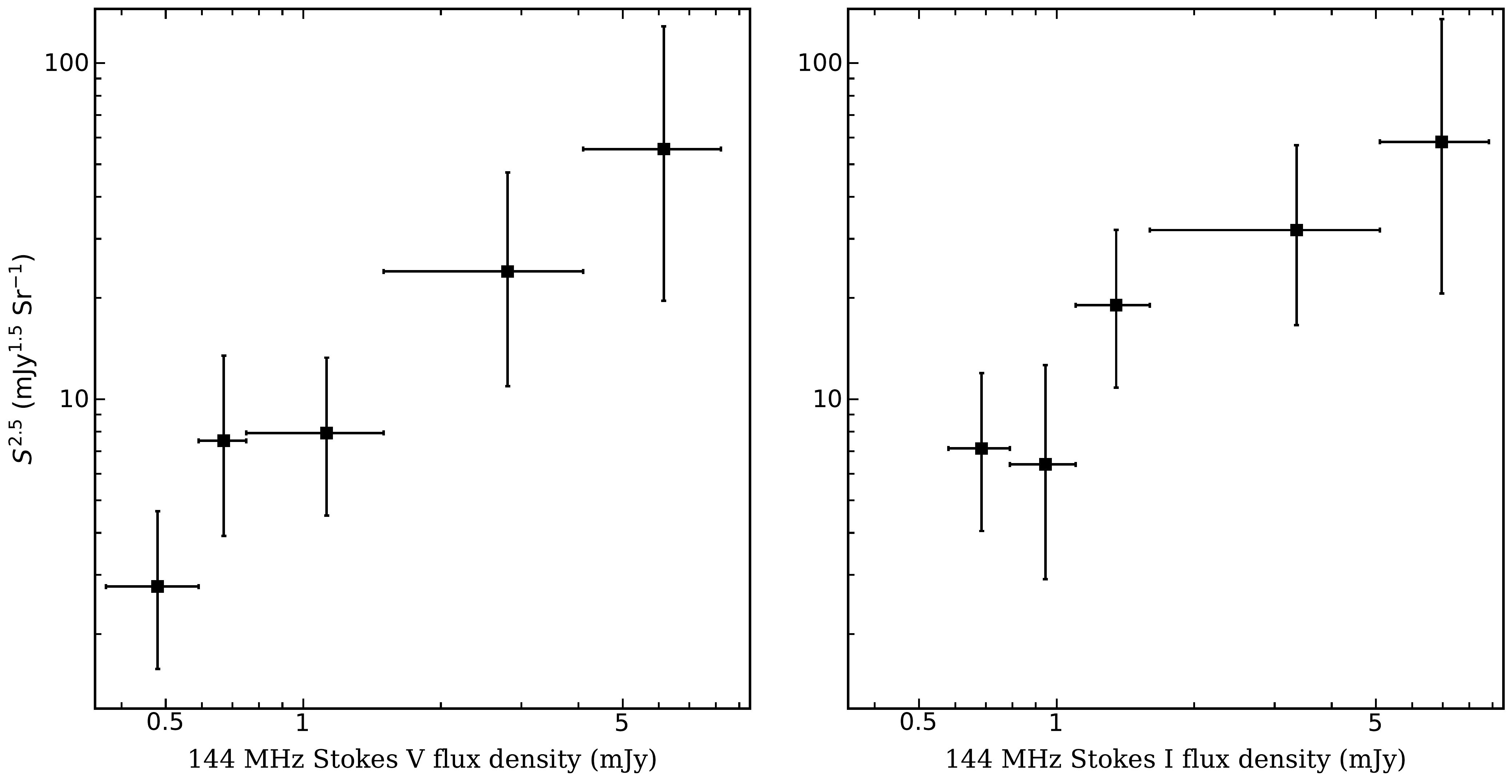}}}
 \caption{{\bf 144\,MHz Stokes~V (left panel) and Stokes~I (right panel) differential source counts for the 17 detected M~dwarf stellar systems.} Since we conducted for the initial source detection in circular polarisation, the Stokes~V source counts are shifted to lower flux densities than the Stokes~I source counts. The error bars represent 1-$\sigma$ uncertainty.} 
\label{fig:source_counts}
\end{figure*}

Supplementary Information Figure\,\ref{fig:source_counts} demonstrates that we have not detected sources with total flux density $S >10\,$mJy in an 8\,hour syntehsised image, and suggests our survey of nearby stellar-systems becomes significantly incomplete at $\approx 400$\,$\mu$Jy. Such a value is consistent with our requirement that a Stokes~V detection be $\geqslant4\sigma$ and the fact the rms noise $\sigma$ of a LoTSS field can vary between $\approx$60 and $\approx$120\,$\mu$Jy. Due to this rms noise variation per LoTSS field, it is also likely that the bin centred at $\approx 700$\,$\mu$Jy is affected by incompleteness.

If we assume that the three bins at $>1$\,mJy are complete, we can estimate how many stellar systems a next-generation survey conducted with Square Kilometre Array (SKA1)-Low will detect. Assuming that the SKA1-Low survey will consist of 8\,hour pointings at 148\,MHz with a bandwidth of $\approx$\,50\,MHz, a $4\sigma$ Stokes~V detection will need to be at least 28\,$\mu$Jy based on current expected array sensitivity\cite{memo,2019MNRAS.484..648P}. We predict that such a survey of the entire Southern sky (declination 0$^{\circ}$ and below) with a Stokes~V detection threshold of 28\,$\mu$Jy will achieve 7900$\pm$2500 detections of M~dwarf systems. Note that this estimate assumes there is no transition in emission mechanism or type of emitter beyond 50\,pc, which appears to be the sensitivity horizon of this work, and no difference in the volume density of detections when surveying the Galactic Plane. If we also include active binaries, such as RS\,CVns, the number of detections will at least double. Finally, the source counts also imply that we will have a total of 100$\pm$30 detections of M~dwarf stellar systems when LoTSS finishes observing and processing the remaining area the Northern sky over the next $\approx$4\,years.

\end{document}